\documentclass[lettersize,journal]{IEEEtran}
\usepackage{amsmath,amsfonts}
\usepackage{algorithmic}
\usepackage{algorithm}
\usepackage{array}
\usepackage[caption=false,font=normalsize,labelfont=sf,textfont=sf]{subfig}
\usepackage{textcomp}
\usepackage{stfloats}
\usepackage{url}
\usepackage{verbatim}
\usepackage{graphicx}
\usepackage{cite}

\graphicspath{{figs/}}  

\usepackage[dvipsnames]{xcolor}
\usepackage{booktabs}
\usepackage{amssymb}
\usepackage{soul}
\usepackage{multirow}
\usepackage{booktabs}
\usepackage{mathtools}
\DeclareMathAlphabet{\mathcal}{OMS}{cmsy}{m}{n}
\usepackage[colorlinks=true, linkcolor=blue, urlcolor=blue, citecolor=blue]{hyperref}
\usepackage{cleveref}

\Crefname{section}{Sec.}{Secs.}
\Crefname{figure}{Fig.}{Figs.}

\crefname{section}{Sec.}{Secs.}
\crefname{figure}{Fig.}{Figs.}

\newcommand{\para}[1]        {\vspace{1pt}\noindent{\textbf{#1}}}
\newcommand{\denselist}{\vspace{-3pt} \itemsep -2pt\parsep=-1pt\partopsep -2pt}

\newcommand {\mm}[1] {\ifmmode{#1}\else{\mbox{\(#1\)}}\fi}

\newcommand{\Xspace}        {\mm{\mathbb{X}}}
\newcommand{\Rspace}        {\mm{\mathbb{R}}}

\newcommand{\etal}{{et al.}}
\newcommand{\eg}{{e.g.,}}
\newcommand{\ie}{{i.e.,}}

\newcommand{\cf}{{cf.}}

\newcommand{\dgm}	     {\mathrm{PD}}
\newcommand{\DTB}               {\mathrm{DTB}}
\newcommand{\minPts}        {\mm{minPts}}
\newcommand{\Ucal}{\mm{\mathcal{U}}}
\newcommand{\Vcal}{\mm{\mathcal{V}}}
\newcommand{\Ccal}{\mm{\mathcal{C}}}
\newcommand{\Rcal}{\mm{\mathcal{R}}}
\newcommand{\Mcal}{\mm{\mathcal{M}}}

\newcommand{\Hgroup}        {\mm{\mathrm{H}}}

\newcommand{\adamapper}	     {AdaMapper}
\newcommand{\adahisomap}	 {AdaHIsomap}
\newcommand{\topomap}	     {TopoMap}
\newcommand{\isomap}	     {Isomap}
\newcommand{\lisomap}	     {L-Isomap}
\newcommand{\hisomap}	     {HIsomap}

\newcommand{\SH}        {\mm{\mathsf{Swiss\,Hole}}}
\newcommand{\VS}        {\mm{\mathsf{Vortex\,Street}}}
\newcommand{\Mice}        {\mm{\mathsf{Mice}}}
\newcommand{\fourelt}        {\mm{\mathsf{4elt}}}
\newcommand{\glasses}        {\mm{\mathsf{Glasses}}}
\newcommand{\fertility}        {\mm{\mathsf{Fertility}}}
\newcommand{\Octa}        {\mm{\mathsf{Octa}}}
\newcommand{\bcsstk}        {\mm{\mathsf{Bcsstk31}}}

\newcommand{\GIF}        {\mm{\mathsf{Cartoon}}}
\newcommand{\nkx}         {\mm{\mathsf{Coauthor}}}
\newcommand{\face}        {\mm{\mathsf{Face}}}
\newcommand{\coauthor}         {\mm{\mathsf{Coauthor}}}

\newcommand{\ct}[1]{\textcolor{ForestGreen}{[cite?]}}

\usepackage{enumitem}
\begin{document}

\title{Homology-Preserving Dimensionality Reduction via Adaptive Mapper and Landmark Isomap}

\author{Shakiba~Khourashahi,
        Ilia~Jahanshahi,
        Bei~Wang,
        Lin~Yan%
\thanks{Shakiba Khourashahi is with Iowa State University. E-mail: shakiba@iastate.edu}%
\thanks{Ilia Jahanshahi is with Iowa State University. E-mail: ilia@iastate.edu}%
\thanks{Bei Wang is with University of Utah. E-mail: beiwang@sci.utah.edu}%
\thanks{Lin Yan is with Iowa State University. E-mail: linyan@iastate.edu}%
}

\maketitle

\begin{abstract}                                                                          
As data becomes increasingly central across engineering and scientific disciplines, effective visualization is essential for interpreting complex, high-dimensional structures. Dimensionality reduction techniques project high-dimensional data into lower dimensions while aiming to preserve structural properties such as pairwise distances and local neighborhoods. In this paper, we focus on improving \emph{homological preservation}, that is, the retention of topological features such as connected components and loops, which is critical for maintaining global shape and continuity. We first introduce {\adamapper}, a Mapper-based algorithm that leverages persistence diagrams to guide both skeleton construction and landmark selection. {\adamapper} incorporates an adaptive refinement strategy that automatically increases cover resolution in regions exhibiting topological loops. 
We then propose {\adahisomap}, which extends landmark Isomap by incorporating homology-informed landmark selection and augmenting it with random anchor points to better balance distance and homology preservation.  
We evaluate both methods on a diverse set of datasets, including high-dimensional point clouds, scientific simulations, networks, and image data, and benchmark their performance against state-of-the-art approaches.
\end{abstract}


\section{Introduction}
High-dimensional datasets are increasingly common across domains such as economics~\cite{StockWatson2002}, biology~\cite{CunninghamSun2025}, physics~\cite{MaljovecWangPascucci2013}, and chemistry~\cite{JoswiakPengCastillo2019}. Visualization provides a critical means to uncover and interpret patterns in such data, with dimensionality reduction (DR) being one of the most widely used approaches.

DR techniques can be broadly grouped into linear and nonlinear methods. 
Linear methods such as principal component analysis (PCA)~\cite{Pearson1901}, multidimensional scaling (MDS)~\cite{torgerson1952}, and Fisher’s linear discriminant (FLD)~\cite{Fisher1936} are valued for their simplicity and employ global linear projections.  
Nonlinear techniques, often associated with manifold learning, capture more complex structures through either metric-based graph constructions (\eg~Isomap~\cite{TenenbaumSilvaLangford2000}, locally linear embedding (LLE)~\cite{RoweisSaul2000}, Laplacian eigenmaps~\cite{BelkinNiyogi2003}) or nonmetric optimizations that preserve dissimilarity rankings~\cite{LiuMaljovecWang2017}.
Beyond this distinction, methods are also categorized by the structural properties they preserve. While distance preservation ensures geometric fidelity (\eg~PCA~\cite{Pearson1901}, MDS~\cite{torgerson1952}), topology preservation emphasizes neighborhood relations (\eg~Isomap~\cite{TenenbaumSilvaLangford2000}, LLE~\cite{RoweisSaul2000}). 
In this paper, we focus on a specific aspect of topological preservation---\emph{homology preservation}---which seeks to preserve 0-dimensional (0D, connected components) and 1-dimensional (1D, loop) features.

Homology-preserving DR is particularly valuable for datasets exhibiting nontrivial topological structure. Persistent homology has been applied across a wide range of domains to analyze high-dimensional point clouds, including biology~\cite{bou2024persistent} and astronomy~\cite{guerrero2024persistent}. For instance, it has been used to identify statistically significant cosmic voids by modeling galaxies as point clouds~\cite{GreenMintzXu2019}.
In image analysis, Carlsson~\etal~\cite{CarlssonIshkhanovDe-Silva2008} and Xia~\etal~\cite{Xia2016} showed that subspaces of natural image patches can exhibit circular structures or even be topologically equivalent to a Klein bottle, depending on patch size. In scientific simulations, topological descriptors such as merge trees and Morse complexes have been used to uncover structural patterns, including clusters, outliers, and periodic behavior. For example, Yan~\etal~\cite{YanMasoodRasheed2022} characterized periodicity in a vortex street dataset by analyzing merging and splitting behavior via merge trees, while Lan~\etal~\cite{LanGamelinYan2024} employed 1-dimensional topological skeletons derived from Morse complexes to study algorithmic uncertainties in atmospheric river detection.
A recent survey~\cite{YanMasoodSridharamurthy2021} provides a broader overview of the integration of homological features into scientific visualization. In complex network science, persistent homology also serves as a fundamental tool for abstraction and summarization, with applications spanning biological and social networks; see~\cite{AktasAkbas2019Fatmaoui} for a comprehensive survey.

Compared with other topology-preserving methods, t-SNE and UMAP have limited control over global structure. The KL (Kullback--Leibler) objective in t-SNE ignores large-distance relationships—points that are far apart in the original space can be placed close together without significant penalty~\cite{MaatenHinton2008}. UMAP~\cite{McinnesHealyMelville2018}, in contrast, optimizes edge-level similarities of a fuzzy $k$-nearest neighbor ($k$NN) graph; however, it does not explicitly preserve end-to-end geodesic distances, and long-range relationships are only indirectly constrained. As a result, both methods may introduce topological artifacts, such as merging or tearing, depending on hyperparameter choices.

In contrast, {\hisomap}, the manifold-landmarking approach of Yan~\etal~\cite{YanZhaoRosen2018}, promotes homology preservation by combining landmark Isomap (L-Isomap) with homology-informed landmark selection. Specifically, it selects landmarks from nodes in the Mapper graph, thereby reducing computational complexity while improving the quality and stability of the resulting embedding.

Despite these advantages, {\hisomap} has two key limitations: (1) sensitivity to parameter choices (its performance depends on five parameters), which can hinder its practical adoption in scientific data analysis, and (2) a globally fixed Mapper resolution that assumes uniform topological complexity along the filter function, often leading to oversampling in non-critical regions and undersampling in regions containing loops. To address these limitations, we propose an adaptive landmarking strategy that extends manifold-landmarking-based DR to more robustly preserve $0$D and $1$D homological features.
Our goal is to provide a general framework for more accurate and insightful data visualization. Our contributions are as follows:
\begin{itemize}[leftmargin=*,noitemsep]
\item \adamapper: A Mapper-based skeletonization method that derives locally adaptive cover resolutions directly from the $1$D persistence diagram, refining loop-critical regions while coarsening non-critical ones. This approach overcomes the limitations of globally fixed Mapper resolutions without extensive manual tuning.
\item {\adahisomap}: Building on {\adamapper}, we propose \adahisomap, which integrates persistence-guided adaptive landmark selection with stochastic anchor points to balance $0$D and $1$D homology preservation. The result is a topology-aware pipeline that leverages homological information to guide intermediate representations for DR.
\item A systematic evaluation across diverse datasets—including high-dimensional point clouds, scientific simulations, networks, and images—demonstrating the behavior of the proposed pipeline under default settings and its competitive performance relative to state-of-the-art DR methods.
\end{itemize}
Our algorithms and datasets is released open source under the MIT license on GitHub (\url{https://github.com/VisTALELab/AdaHIsomap.git}). 
\section{Related Work}
\label{sec:related}
We review related work in two areas most relevant to our contributions: topology-aware skeletonization methods based on Mapper and structure-preserving DR techniques.

\para{Mapper-based data skeletonization.}
The Mapper construction~\cite{SinghMemoliCarlsson2007} captures the topological structure of a high-dimensional point cloud and summarizes it as a $1$D skeleton graph. It has been widely adopted in topological data analysis and visualization, with applications spanning finance~\cite{ShirajRahman2024} and medical imaging~\cite{HaseganGeniesse2024,LoughreyFitzpatrick2021}.

A central challenge of Mapper is its sensitivity to the choice of filter function, cover parameters, and clustering parameters. These user-specified parameters can significantly influence the resulting graph, complicating both interpretation and reproducibility. Motivated by its broad applicability, numerous variants have been proposed to improve robustness and extend Mapper to new domains~\cite{OulhajCarriereMichel2024,ChalapathiZhouWang2021,BUIBayVo2020,AlvaradoBeltonFischer2025,scoccolalimHarrington2025}.

Soft Mapper~\cite{OulhajCarriereMichel2024} addresses sensitivity to the filter function by introducing an optimization framework that selects filters automatically based on topological criteria. Other approaches focus on cover parameter selection. For example, multi-pass AIC/BIC~\cite{ChalapathiZhouWang2021}, inspired by the X-means algorithm~\cite{pellegMoore2000} (an extension of K-means~\cite{Lloyd1982}), constructs adaptive covers using information-theoretic model selection. F-Mapper~\cite{BUIBayVo2020} employs fuzzy C-means clustering~\cite{bezdek1981} to allow overlapping clusters. More recently, Alvarado~\etal~proposed G-Mapper~\cite{AlvaradoBeltonFischer2025}, which leverages G-means clustering with Gaussian mixture models (GMMs) and iterative splitting to eliminate the need for a priori interval counts, adapting cover elements to the underlying data distribution. Scoccola \etal~\cite{scoccolalimHarrington2025} further advanced this direction by introducing a general cover-learning framework that adapts automatically to data-driven topological features.

These methods substantially improve the usability and stability of Mapper by automating filter or cover selection. However, they do not explicitly leverage persistent homology to control cover resolution locally along the filter function.

Compared with prior Mapper variants, {\adamapper} (detailed in~\cref{sec:AdaMapper}) derives key cover parameters directly from properties of $1$D persistent homology, reducing reliance on user expertise while still allowing optional user control. Specifically, it introduces a locally adaptive cover that increases resolution in regions exhibiting $1$D loops while avoiding unnecessary refinement elsewhere. As a result, {\adamapper} produces more stable and informative Mapper skeletons and is more robust across datasets with heterogeneous topological structure.

\para{Structure-preserving dimensionality reduction.}
Structure-preserving DR aims to recover intrinsic shapes in data, such as manifolds, spirals, or curved surfaces. UMAP~\cite{McinnesHealyMelville2018} constructs a fuzzy $k$NN graph and learns a low-dimensional embedding that preserves local structure while approximating global relationships. t-SNE~\cite{MaatenHinton2008} is a probabilistic DR method that models local pairwise similarities and emphasizes neighborhood preservation, producing embeddings that are stable and visually interpretable.
Another notable approach is Isomap~\cite{TenenbaumSilvaLangford2000}, which captures manifold geometry using geodesic distances, that is, the shortest paths along the underlying manifold. This is particularly effective for data lying on nonlinear manifolds, where Euclidean distances can underestimate true separations between distant points and overestimate distances between points that are close along the manifold but far in ambient space. Landmark Isomap (L-Isomap)~\cite{SilvaTenenbaum2002} addresses the computational cost of Isomap by selecting a subset of points as landmarks to approximate the distance matrix, thereby reducing complexity while largely preserving geometric fidelity.

More recently, DR techniques have been developed with an explicit focus on preserving topological features. TopoMap~\cite{DoraiswamyTiernySilva2020} preserves the $0$D persistence diagram of the Vietoris--Rips filtration during embedding by leveraging the Euclidean minimum spanning tree and iteratively merging connected components to maintain topological structure. Building on this approach, TopoMap++~\cite{GuardieiroOliveiraDoraiswamy2024} improves computational efficiency and introduces an interactive framework for exploring the hierarchical organization of topological components. 
In parallel, a growing body of work incorporates topology-aware loss functions into machine learning models for homology-preserving DR. For example, TopoAE~\cite{MoorHornRieck2020} introduces a topological loss based on discrepancies in $0$D homology between the original and embedded spaces. This framework was recently extended to preserve $1$D persistent homology by Cl\'emot~\etal~\cite{ClemotDigneTierny2025}, who proposed TopoAE++, a cycle-aware topological autoencoder.

The work most closely related to ours is that of Yan~\etal~\cite{YanZhaoRosen2018}, which is designed to preserve both $0$D and $1$D homological features. Referred to as {\hisomap} in this paper, the method builds on the landmark selection strategy of L-Isomap, using a data skeleton derived from the Mapper graph~\cite{ShinagawaKuniiKergosien1991} to guide the selection of representative landmarks on the manifold.

In this work, we adopt the landmark Isomap paradigm but reformulate the landmark selection strategy and associated pipeline components. Specifically, we use {\adamapper} to derive landmarks from a persistence-guided, adaptively constructed skeleton, and we introduce random anchor points to balance distance preservation with $0$D and $1$D homology preservation in regions not associated with prominent loops. The resulting DR pipeline is evaluated against the structure-preserving methods discussed above in~\Cref{sec:results}.

\section{Technical Background}
\label{sec:background}
We review persistent homology, Mapper, and~\hisomap, which form the foundations of our proposed methods. We then describe how these techniques are customized to develop {\adamapper} (\cref{sec:AdaMapper}) and {\adahisomap}  (\cref{sec:AdaHIsomap}).

\begin{figure*}[t]
\centering 
\includegraphics[width=1.8\columnwidth]{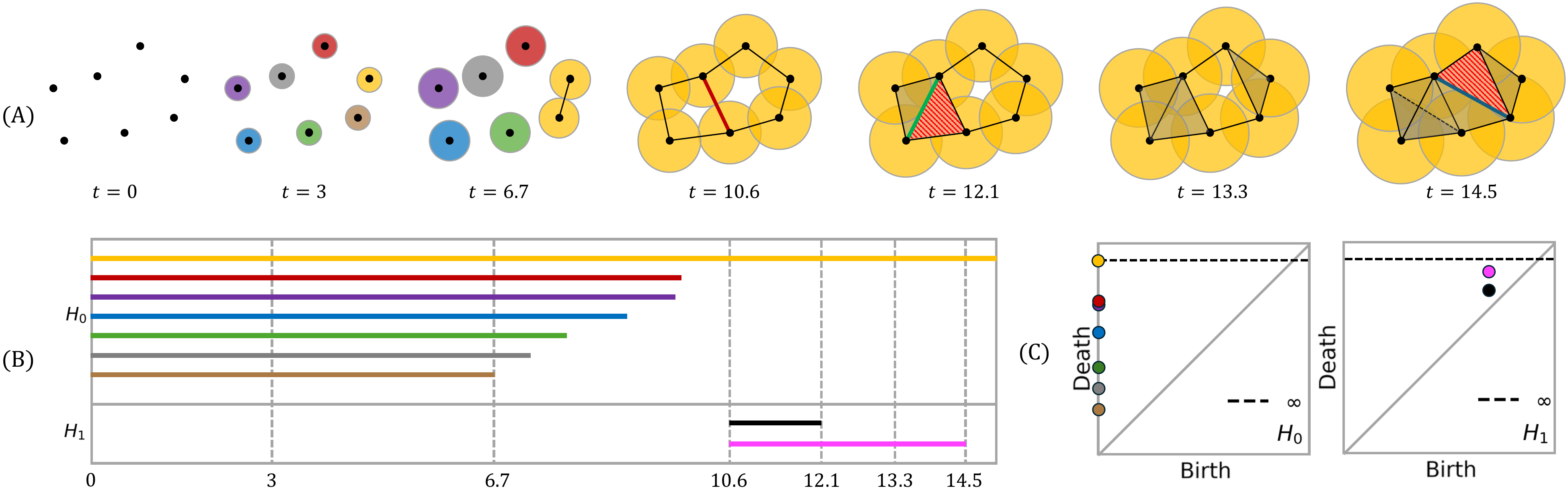}
\vspace{-2mm}
\caption{Computing persistent homology of a point cloud. (A) Simplicial complexes constructed on seven points at increasing values of $t$. (B) Corresponding persistence barcodes. (C) Persistence diagrams for 0D (left) and 1D (right) features.}
\label{fig:PD}
\vspace{-2mm}
\end{figure*}

\subsection{Persistent Homology}
\label{sec:PH}
Homology provides a formal framework for distinguishing topological spaces based on their structural features, particularly the presence of ``holes'' across various dimensions. Mathematically, homology associates a topological space with a sequence of abelian groups, known as homology groups, which encode the number and type of topological features present in the space~\cite{Ghrist2008}. These groups capture essential information about the space’s connectivity. Given a topological space $\Xspace$, its homology groups $\Hgroup_0(\Xspace)$, $\Hgroup_1(\Xspace)$ and $\Hgroup_2(\Xspace)$ capture structural features of increasing dimension: $\Hgroup_0(\Xspace)$ represents connected components, $\Hgroup_1(\Xspace)$ represents loops or tunnels, and $\Hgroup_2(\Xspace)$ represents voids. By systematically quantifying these features, homology enables the comparison of topological spaces through their algebraic invariants. This characterization is foundational to the persistent homology framework, which extends classical homology to study how topological features evolve across different scales.

To compute homology from discrete data such as point clouds, we approximate the underlying space with a simplicial complex, a combinatorial structure built from vertices, edges, triangles, and higher-dimensional simplices that generalizes the notion of meshes; see the black skeleton in~\cref{fig:PD}(A) for an example. A simplicial complex provides a finite, structured representation of the space, enabling efficient algebraic operations for computing homology groups.

As illustrated in~\cref{fig:PD}, we can now construct a filtration of nested simplicial complexes to study homology  across scales.
Given a distance metric $\mathcal{D}_X$ (\eg~the Euclidean metric), we construct a sequence of nested topological spaces by forming unions of balls of radius $t$ centered at each point in $X$, where $t$ varies over a non-negative scale. 
Persistent homology then records the birth and death of topological features within this filtration, capturing their evolution across scales.

In~\cref{fig:PD}(A), at $t = 0$, each point is \emph{born} (appears) as its own (connected) component. As $t$ increases, we focus on the important events when components merge with one another to form larger components or tunnels. 
We begin by tracking the birth and death times of each component or tunnel as well as its lifetime in the filtration.
At $t = 6.7$, the brown component merges into the orange component and \emph{dies} (disappears); therefore, the brown component has a lifetime (i.e., \emph{persistence}) of $6.7$.
At $t = 7.3$, the gray component merges into the purple component and dies; therefore, it has a persistence of $7.3$.
So on and so forth. 
At $t = 10.6$, the collection of components forms two tunnels; see two loops beside the red edge in~\cref{fig:PD}(A). This edge is the birth simplex responsible for the creation of the loops.

The left tunnel is filled (disappears/dies) at $t=12.1$ with the appearance of the red triangle with a green edge, whereas the right tunnel is filled (disappears/dies) at $t=14.5$ with the appearance of the red triangle with a blue edge.
These simplices, responsible for the death of loops, are called the \emph{death triangles}
in~\cref{sec:criticalRange}.
The orange component born at time $0$ never dies, therefore, it has a persistence of $\infty$.
We record and visualize the appearance (birth), the disappearance (death), and the persistence of homological features in the filtration via persistence diagrams~\cite{Cohen-SteinerEdelsbrunnerHarer2007} (\cref{fig:PD}(C)), or equivalently, persistence \emph{barcodes}~\cite{Ghrist2008b} (\cref{fig:PD}(B)). 
A point $\rho=(b,d)$ in the persistent diagram of $X$ records a homological feature that is born at time $b$ and dies at time $d$. 0D and 1D persistence diagrams, denoted as $\dgm_0(X)$ and $\dgm_1(X)$, capture the births and deaths of components and tunnels.  
Equivalently in the barcode of~\cref{fig:PD}(B), such a feature is summarized by a horizontal bar that begins at $b$ and ends at $d$.

From a computational perspective, a filtration can be represented as a sequence of simplicial complexes, providing a compact and algorithmically tractable encoding (see~\cite{EdelsbrunnerHarer2010} for details). 
Moreover, birth and death simplices offer spatial clues to the location of the corresponding topological structures~\cite{Okediji2024}. 

In this paper, persistent homology is used to extract structural information, rather than to recover explicit cycle representatives. 
It is used in~\cref{sec:AdaMapper} to guide the construction of a Mapper that subsequently supports the preservation of 0D and 1D features in lower-dimensional space in~\cref{sec:AdaHIsomap}.

\begin{figure}[t]
 \centering 
 \includegraphics[width=0.98\columnwidth]{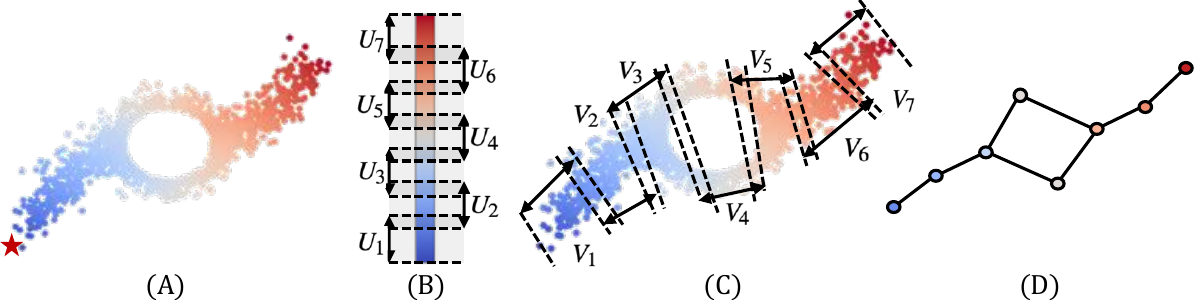}
 \vspace{-2mm}
 \caption{Illustration of the Mapper pipeline. (A) Point cloud $X$, colored by the DTB function $f(X)$, with the base point marked by a red star. (B) Cover $\Ucal$ of $f(X)$ using open intervals with $m=7$ and overlap $p=0.2$. (C) Induced cover $\Vcal$ of $X$, formed by clustering points in preimages $f^{-1}(a_i, b_i)$ for each $i$. (D) Mapper construction, given by the nerve of $\Vcal$.}
 \label{fig:mapper}
  \vspace{-4mm}
\end{figure}

\subsection{Mapper}
\label{sec:mapper}
Our method adopts a topological approach to data summarization, using the Mapper algorithm~\cite{SinghMemoliCarlsson2007} to guide the selection of landmarks for dimensionality reduction. To support this, we briefly review the definitions and computational steps underlying Mapper, which yields a discrete representation of the data’s structural features.

\textbf{Filter function.}
Given a topological space $\Xspace$, a continuous function $f: \Xspace \to \Rspace$, referred to as the \emph{filter function}, plays a critical role in determining how structural features of the data are revealed. In practice, the filter function is chosen to encode geometric or topological information relevant to the analysis task~\cite{BiasottiGiorgiSpagnuolo2008}. Common examples include height functions, barycentric distance, curvature, and geodesic distances. In this work, we use the geodesic distance from a fixed source point as the filter function, referred to as the distance-to-basepoint ($\DTB$) function. 
Prior studies have shown that such functions are effective in revealing 1D homological structures, including prominent loops and circular patterns~\cite{BiasottiGiorgiSpagnuolo2008, GeSafaBelkin2011}.
While DTB is used as the default in our experiments, our framework is not restricted to this choice; see a discussion in~\cref{sec:limAdaMapper}.

\textbf{Mapper graph.}
To handle discrete point cloud data, we utilize the Mapper algorithm~\cite{SinghMemoliCarlsson2007}, which builds a graph-like summary called a skeleton by analyzing the behavior of the filter function over overlapping regions of the domain.
Mapper partitions the image of the filter function into overlapping intervals and examines the pre-images of these intervals to identify structural patterns in the data~\cite{SaulVeen2017}.

As illustrated in~\cref{fig:mapper}, given a point cloud $X$ and a filter function $f$, we construct a cover $\mathcal{U}$ of $f(X)$ using overlapping intervals controlled by the number of intervals $m$ and overlap ratio $p$. For example, Fig.~\ref{fig:mapper}(B) shows the $\DTB$ function partitioned into $m=7$ intervals with a $0.2$ overlap, forming $\mathcal{U} = [U_1, \ldots, U_7]$. Each interval $U_i$ induces a pre-image $V_i = f^{-1}(U_i)$, and the collection $\mathcal{V} = \{V_i\}$ forms a cover of $X$. Each $V_i$ is then clustered to identify components.
In our implementation, we use DBSCAN~\cite{EsterKriegelSander1996}, a density-based clustering method that can identify arbitrarily shaped clusters without requiring the number of clusters to be specified in advance. Other clustering strategies could be used without altering the overall Mapper pipeline.
The resulting clusters form the nodes of the Mapper graph, with edges added between nodes that share data points, as illustrated in~\cref{fig:mapper}(C).

The resulting structure, denoted $\mathcal{M}$, is referred to throughout this paper as the \emph{homological skeleton}. A depiction of the full Mapper output is shown in \cref{fig:mapper}(D), illustrating how the data’s topological structure is encoded in a compact, interpretable graph with branches and loops.

\subsection{\hisomap}
\label{sec:isomap}
In the context of non-linear DR, methods such as Isomap, L-Isomap, and the more recent~\hisomap~\cite{YanZhaoRosen2018}~play a significant role in the preservation of the intrinsic geometry of high-dimensional data. 
Give a point cloud $X$, the implementation of~\hisomap~involves four steps:

\para{Step 1: $k$NN graph construction.} Build the $k$NN graph $G$ on $X$.

\para{Step 2: Filter function.} Compute the DTB filter function \(f: X \to \mathbb{R}\), which encodes structural properties relevant to DR. 
The base point is chosen from the extreme points of 
 \(X\).

\para{Step 3: Mapper and landmark selection.} Compute Mapper graph $\Mcal$ with user-specified resolution parameters. Nodes of \(\Mcal\) correspond to clusters of points in \(X\); the cluster centroids are selected as the landmarks denoted as $X_L \subseteq \Xspace$. 

\para{Step 4: Apply L-Isomap}. Replace the randomly generated landmarks in L-Isomap with $X_L$, and apply the rest of the L-Isomap algorithm.

For clarity, we distinguish different versions of L-Isomap according to their landmark selection schemes: the variant with randomly selected landmarks is referred to as~\emph{random L-Isomap}, while the homology-preserving approach from~\cite{YanZhaoRosen2018}~is denoted as {\hisomap}. In this paper, \hisomap~serves as a conceptual and algorithmic baseline for the topology-aware DR pipeline developed in subsequent sections.

\section{{\adamapper}: A Mapper-based algorithm for homological skeleton extraction}
\label{sec:AdaMapper}

The core idea of {\adamapper} is to use persistent homology in the original space to guide Mapper construction in a homology-informed manner. In this section, we present the technical details.

\begin{figure}[t]
\centering 
\includegraphics[width=\columnwidth]{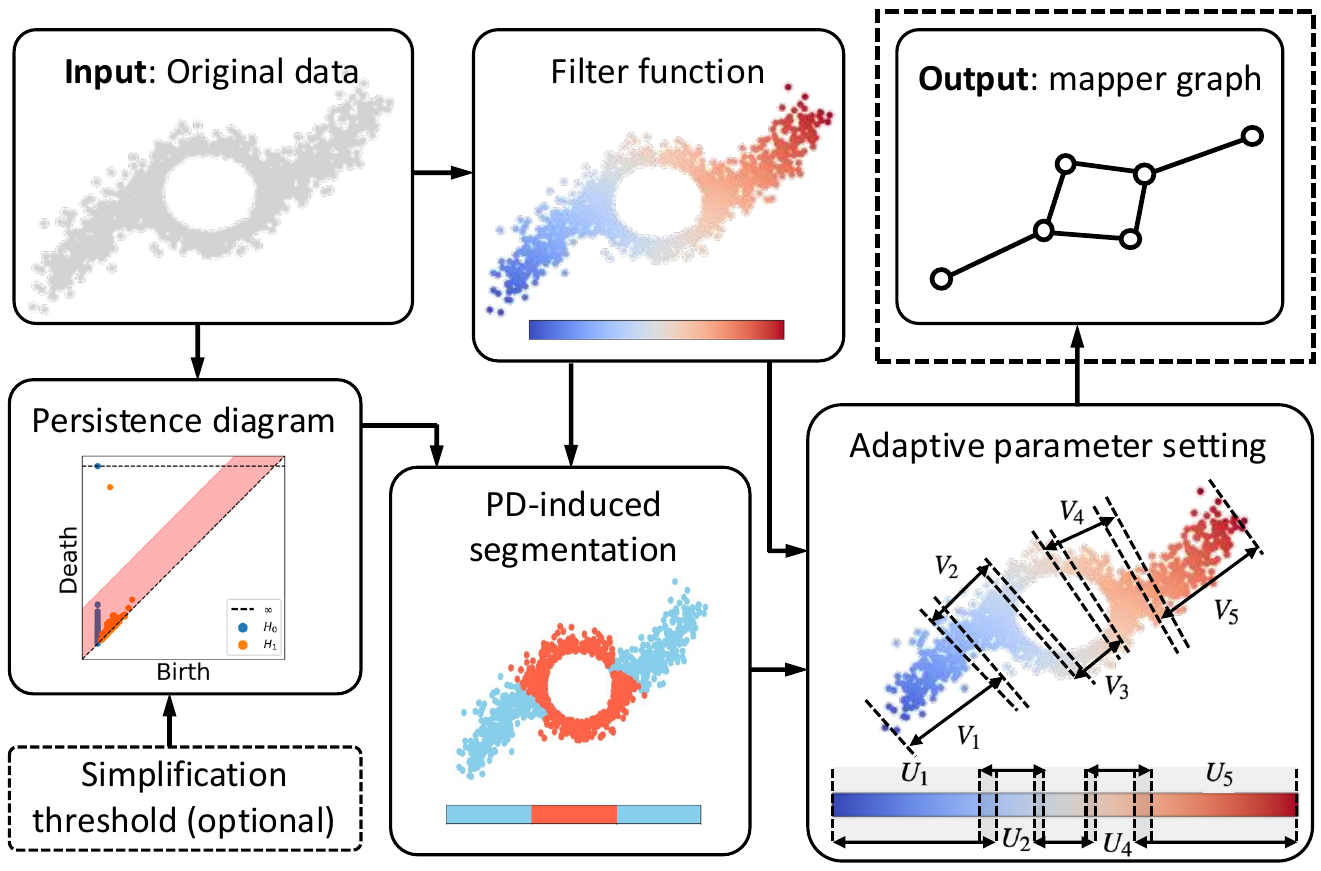}
\vspace{-6mm}
\caption{{\adamapper} pipeline for homology-informed Mapper construction.}
\vspace{-4mm}
\label{fig:AdaMapper}
\end{figure}

\para{Overview.} 
An overview of~{\adamapper} is shown in~\cref{fig:AdaMapper}. As a preliminary step, we construct the $k$NN graph on the input data. This graph is used both to compute the DTB filter function and by \adahisomap, so we build it once upfront. The main procedure then proceeds as follows. First, we compute the persistence diagram of the input data to characterize its topological structure. 
Second, we compute the filter function and partition its range into regular and critical segments, guided by the significant 1D features (loops) identified in the persistence diagram (\cref{sec:criticalRange}). Third, we derive parameters for the Mapper construction separately for the regular and critical segments, based on the persistence properties of loops (\cref{sec:internalParam}). 
We define “significant” loops according to their persistence, using a default threshold of $0.35$ (\ie~35\% of the maximum persistence). 
Users may also adjust this threshold interactively, depending on their interpretation of the persistence diagram (\cref{sec:smpf}).

\para{Notations.} Let $X$ denote the original data and $f: X \to \Rspace$ its filter function. We use the 1D persistence diagram $\dgm_1(X)$ to separate $f$ into the critical segment and the regular segment, denoted $\Ccal$ and $\Rcal$, respectively. With PD-induced segmentation, we calculate the cover $\Ucal$ of $f(X)$ and then derive the cover $\Vcal$ of $X$. 
The Mapper graph $\Mcal$ is the 1D skeleton of the nerve of $\Vcal$.
The input to {\adamapper} is $X$ and the output is $\Mcal$. 

With its default settings, {\adamapper} requires no user-specified parameters. However, users may optionally incorporate prior knowledge by specifying parameters such as the persistence threshold $T$, the overlap percentage between adjacent intervals $p$, and the clustering algorithm settings. If $\dgm_1(X)$ does not exhibit any significant loops, the user is prompted to specify the number of intervals $m$, in which case a standard Mapper is applied.

\subsection{PD-Induced Segmentation}
\label{sec:criticalRange}
The first step of {\adamapper} is to compute the persistence diagram of the original data. This persistence diagram, $\dgm_1(X)$, captures the underlying topological structure of the data. The key idea behind {\adamapper} is to leverage the homological information encoded in $\dgm_1(X)$ to identify a critical segment $\Ccal$ along $f$, where finer-grained intervals are required to preserve the selected loops in the Mapper construction.

Let $\rho_i \in \dgm_1(X)$ be a loop whose persistence exceeds the threshold $T$ (with $T=0.35$). For each such loop, we determine its corresponding range on $f$, denoted by $r_i=[u_i,v_i]$. The \emph{critical segment} $\Ccal$ of $f$ is then defined as the union of all such ranges,
$
\Ccal = r_1 \cup r_2 \cup \cdots \cup r_M,
$
where $M$ is the number of loops whose persistence is greater than the simplification threshold. The remaining portion of $f$ is referred to as the \emph{regular segment}, denoted by $\Rcal$. Thus,
$
f = \Ccal \cup \Rcal.
$
We refer to this procedure as the \emph{PD-induced segmentation} of $f$. In the remainder of this subsection, we describe how to determine $r_i$ given $X$, $\rho_i$, and $f$.
 
\subsubsection{Death-Triangle-Based Localization} 
For each loop $\rho_i$, we identify its associated death triangle, which corresponds to the 2-simplex responsible for the disappearance of the loop in the filtration (\cref{sec:PH}). This death triangle provides the basis for determining the loop-critical interval $r_i$.

\begin{figure}[t]
\centering 
\includegraphics[width=\columnwidth]{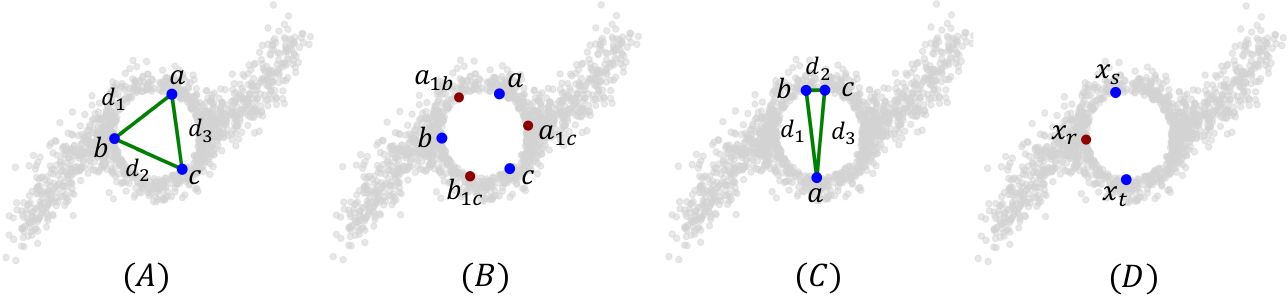}
\vspace{-6mm}
\caption{
(A) Balanced death triangle.
(B) Centroids of the shortest paths between vertex pairs $(a,b)$, $(b,c)$, and $(a,c)$, denoted by $a_{1b}$, $b_{1c}$, and $a_{1c}$, respectively.
(C) Unbalanced death triangle.
(D) Vertices $x_s$ and $x_t$ are identified from the longest edge of the death triangle, and the centroid $x_r$ is computed along their shortest path.
}
\vspace{-2mm}
\label{fig:deathEdge}
\end{figure}

\para{\underline{Step 1: Identify triangle shape.}} 
Given a death triangle (calculated by Ripserer.jl library~\cite{Cufar2020,CufarVirk2023}), let its vertices be $a$, $b$, and $c$. 
We calculate the Euclidean distances between each pair of these vertices in the original point cloud: $d_1$ for $(a,b)$, $d_2$ for $(b,c)$, and $d_3$ for $(a,c)$.
To assess the uniformity of the triangle, we compute the normalized entropy
$H = -\sum_{i=1}^{3} p_i \frac{\log(p_i)}{\log(3)}$, where $p_i = \frac{d_i}{d_1 + d_2 + d_3}$.

A higher entropy value indicates more uniformly distributed edge lengths. 
Based on empirical observation, if $0.975 \leq H \leq 1$, the death triangle is classified as \textit{balanced} (\cref{fig:deathEdge}(A)); otherwise, it is classified as \textit{unbalanced} (\cref{fig:deathEdge}(C)).

\para{\underline{Step 2a:~Derive $r_i$ for balanced death triangle.}} 
Using the $k$NN graph $G$ (already constructed for computing the filter function), we compute shortest paths between each pair of vertices $(a,b)$, $(b,c)$, and $(a,c)$, and denote the centroids of these paths as $a_{1b}$, $b_{1c}$, and $a_{1c}$ (see \cref{fig:deathEdge}(B)).
The interval $r_i$ is then defined by the minimum and maximum filter values among the death-triangle vertices and the three centroids.
$
r_i = [\min f(S), \; \max f(S)]$, 
where $S = \{a, b, c, a_{1b}, b_{1c}, a_{1c}\}$.

\para{\underline{Step 2b:~Derive $r_i$ for unbalanced death triangle.}} 
If the death triangle edges are not balanced (\cref{fig:deathEdge}(C)), we identify the longest edge of the triangle and denote its two vertices as \(x_s\) and \(x_t\). 
Using these vertices and the $k$NN graph \(G\), we compute the shortest path between \(x_s\) and \(x_t\). The centroid of this path is then chosen as \(x_r\) (\cref{fig:deathEdge}(D)).

\begin{figure}[t]
\centering 
\includegraphics[width=.9\columnwidth]{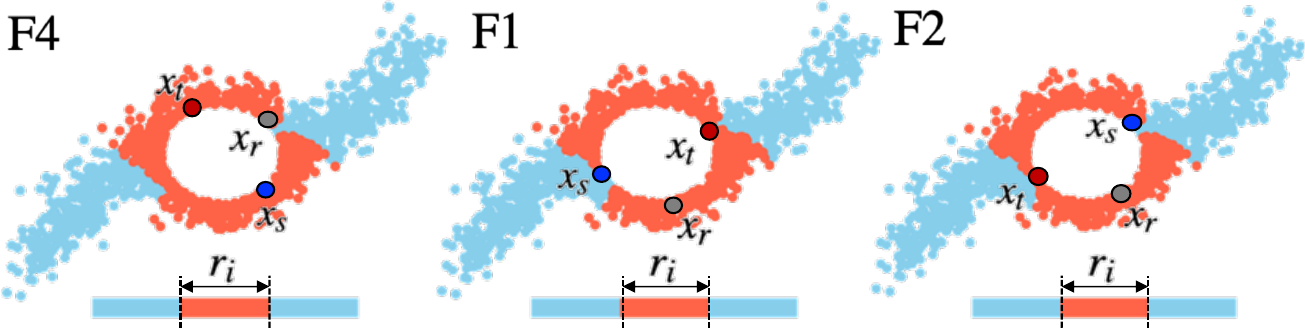}
\vspace{-2mm}
\caption{Derivation of $r_i$ from $x_s$, $x_t$, and $x_r$ under cases F4, F1, and F2.}
\vspace{-2mm}
\label{fig:ri}
\end{figure}
We can now derive $r_i$ considering the following four cases.
\begin{itemize}[noitemsep]
    \item[F1:] If $f(x_s) \leq f(x_r) \leq f(x_t)$, then
    $r_i = [f(x_s),\, f(x_t)]$.

    \item[F2:] If $f(x_t) \leq f(x_r) \leq f(x_s)$, then
    $r_i = [f(x_t),\, f(x_s)]$.

    \item[F3:] If $f(x_r) \leq \min\{f(x_s), f(x_t)\}$, then
    $r_i = [f(x_r),\, f(x_s) + (f(x_t) - f(x_r))]$.

    \item[F4:] If $f(x_r) \geq \max\{f(x_s), f(x_t)\}$, then
    $r_i = [f(x_s) - (f(x_r) - f(x_t)),\, f(x_r)]$.
\end{itemize}

\Cref{fig:ri} illustrates the derivation of $r_i$ under three representative cases. From left to right, the panels correspond to cases F4, F1, and F2. In each case, the interval $r_i=[u_i,v_i]$ derived from $x_s$, $x_t$, and $x_r$ is highlighted in red along the filter function $f$, together with the points whose filter values fall within this interval. Since the dataset contains a single loop, the critical segment is $\Ccal=r_i$, and the complement defines the regular segment $\Rcal$, shown in blue.
Not all intervals $r_i$ in \cref{fig:ri} fully cover the points surrounding the feature $\rho_i$, and this issue is exacerbated when $\rho_i$ has an irregular shape. However, a suboptimal choice of $r_i$ does not necessarily prevent the construction of a cover $\Vcal$ of $X$ that captures $\rho_i$; see~\cref{sec:impectr} for further discussion.

After computing all \(r_i\) (for $i\in [1,M]$), the function \(f\) is partitioned into critical and non-critical regions, from which we derive adaptive parameters to construct the Mapper cover, as detailed in~\cref{sec:internalParam}.

\subsubsection{Design Rationale and Tradeoffs}
The goal of the death-triangle strategy is not to recover a geometrically minimal or complete set of points enclosing a loop, but rather to provide a reliable localization of loop-critical regions suitable for {\adamapper} construction. This design reflects the role of PD-induced segmentation in {\adamapper}, where loop localization is used to control Mapper resolution rather than to extract explicit cycle representatives.

Several existing approaches aim to identify a subset of $X$ that geometrically surrounds a loop represented in $\dgm_1(X)$, including representative cycles via involuted homology~\cite{CufarVirk2023}, reconstructed shortest cycles~\cite{Cufar2020}, and linear-programming-based optimal cycle representatives~\cite{LiThompsonHenselman2021}, among others~\cite{Saul2018}. While effective for explicit loop reconstruction, these methods can become unstable when multiple loops are spatially proximate or interacting, causing the associated regions to expand uncontrollably or mix contributions from multiple features.

By contrast, the death-triangle strategy prioritizes reliable localization over geometric tightness. By relying on the simplices responsible for the disappearance of loops in the filtration, it yields localizations that are consistent with persistence pairings and robust to interactions among nearby features. As a result, the derived interval $r_i$ may not fully enclose the geometric extent of a loop, particularly for irregular or geometrically complex features.

Importantly, this conservative localization is intentional. 
In the context of {\adamapper} construction, recovering a compact and stable loop-critical region is often more effective than including all points surrounding a loop. 
Overly aggressive localization that includes substantially more points than necessary can be counterproductive for {\adamapper} parameterization, a phenomenon analyzed in detail in~\cref{sec:impectr}. 
Nevertheless, although $r_i$ may be imperfect, it provides sufficient information to guide the downstream construction and preserve the presence of the corresponding feature $\rho_i$.

\subsection{{\adamapper} Parameterization}
\label{sec:internalParam}

\begin{figure}[t]
\centering 
\includegraphics[width=0.85\columnwidth]{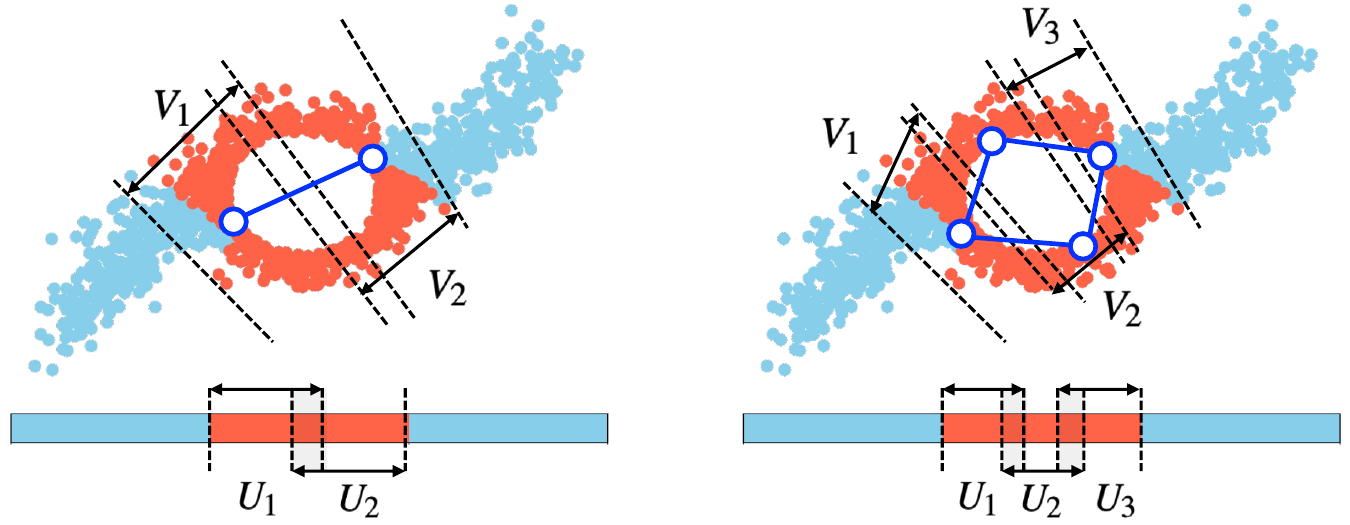}
\vspace{-2mm}
\caption{Critical segment for $m = 2$ (left) and $m = 3$ (right).}
\label{fig:criticalRangeInti}
\end{figure}

The derivation of internal parameters for {\adamapper} starts from the observation that a Mapper requires a cover with at least three elements ($\{V_i\}_{i=1}^{m}$, $m\ge 3$) to identify a loop using the point cloud around it. 
Using data points only within the critical segment, \cref{fig:criticalRangeInti} demonstrates the Mappers constructed with different sizes of cover.
The Mapper captures the loop with $m = 3$, while it fails to summarize the loop with $m = 2$; \cf~\cref{fig:criticalRangeInti}(right) and~\cref{fig:criticalRangeInti}(left).

\begin{figure}[t]
\centering 
\includegraphics[width=0.9\columnwidth]{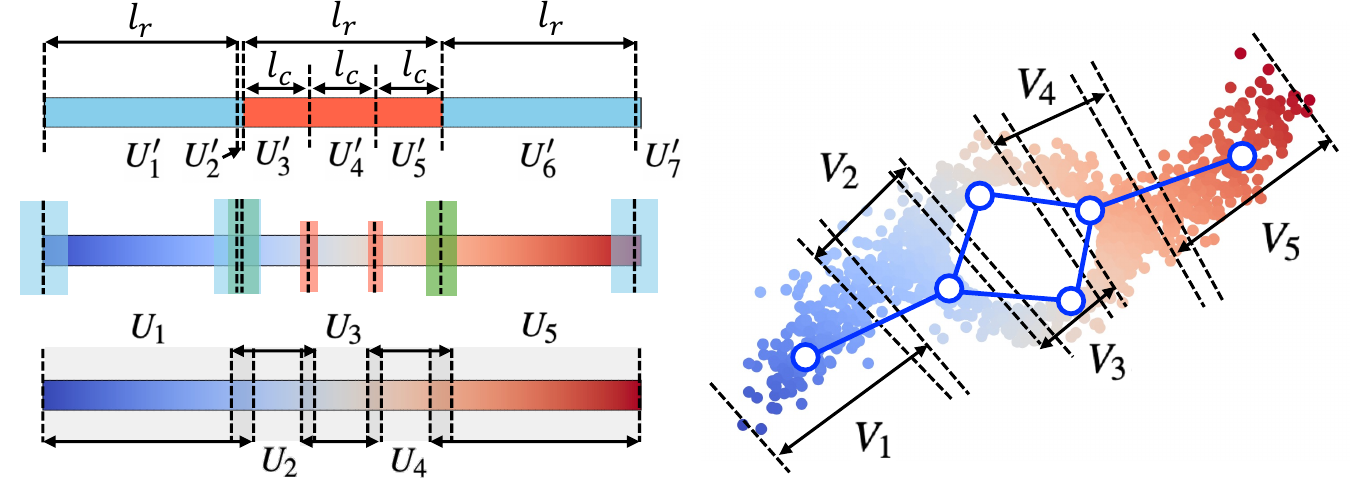}
\vspace{-2mm}
\caption{Adaptive parameter setting for {\adamapper}. Left: three-step construction of open intervals $\Ucal$. Right: the corresponding cover $\Vcal$ of $X$ and the resulting Mapper.}
\vspace{-2mm}
\label{fig:Adamapper-PmSetting}
\end{figure}

\subsubsection{Parameter Derivation for {\adamapper} Construction}
Motivated by this observation, we derive the cover $\Ucal$ of $f(X)$ and then the cover $\Vcal$ of $X$ in the following three steps.

\para{\underline{Step 1:~Initialize the length of open interval.}} 
Among the loops selected by the simplification threshold, denoted by $\rho_1,\dots,\rho_M$, let the lengths of their corresponding ranges be $|r_i| = |v_i-u_i|$ for $i=1,\dots,M$. We define the shortest range by
$
|r_{\min}| := \min\{|r_1|,\dots,|r_M|\}.
$
The initial length of open intervals within the critical segment $\mathcal{C}$ is set to
$
l_c = |r_{\min}|/3,
$
while the initial length of open intervals within the regular segment $\mathcal{R}$ is set to
$
l_r = |r_{\min}|.
$

Based on these settings, we construct an initial cover $\mathcal{U}'$ of $f(X)$, where
$
\mathcal{U}' = \{U'_i\}_{i=1}^{m'} = \{[a'_i,b'_i]\}_{i=1}^{m'},
$
with adjacent intervals satisfying $b'_i = a'_{i+1}$. Furthermore,
$
|a'_i-b'_i| = l_c
$
if $U'_i \in \mathcal{C}$, and
$
|a'_i-b'_i| = l_r
$
if $U'_i \in \mathcal{R}$. 
Note that the initial number of intervals $m'$ may differ from the final number of intervals $m$ after the following two refinement steps.

Using the PD-induced segmentation shown in~\cref{fig:criticalRangeInti}, we observe that this dataset contains only one significant loop, and $|r_{\min}|$ corresponds to the length of the red region. Based on this observation, the initialization of $l_c$ and $l_r$ is illustrated at the top of~\cref{fig:Adamapper-PmSetting}(left). We further derive the initial cover $\mathcal{U}'$ of $f(X)$ with $m' = 7$.
Specifically, we set $l_c = |r_{\min}|/3$ according to the structure observed in~\cref{fig:criticalRangeInti}. In contrast, we choose $l_r = |r_{\min}|$, since regions without a prominent loop are expected to exhibit fewer structural changes.

The latter choice is also a natural implementation of persistence simplification on Mapper, which will be discussed in~\cref{sec:smpf}.

\para{\underline{Step 2:~Adaptive expand $l_c$ and $l_r$ with $p$}}. 
The classic Mapper algorithm uses a parameter $p$ to specify the percentage of overlap between each pair of adjacent intervals, thereby enforcing a uniform overlap across the entire cover. In {\adamapper}, we instead adaptively expand the interval lengths $l_c$ and $l_r$ to generate overlaps of varying sizes, using a default overlap ratio of $p=0.2$.

These expansions may occur in three different settings: between two intervals in the critical segment (C1), between two intervals in the regular segment (C2), and between one interval in the critical segment and one in the regular segment (C3). We design separate expansion strategies for each case. 

Given $U'_i=[a'_i, b'_i]$ and $U'_{i+1}=[a'_{i+1}, b'_{i+1}]$ with $b'_i=a'_{i+1}$, 

\begin{itemize}\denselist
\item[C1:] \underline{if $U'_i\in \Ccal$ and $U'_{i+1}\in \Ccal$}, we expand $U'_i$ from $[a'_i, b'_i]$ to $[a'_i, b'_i+p/2\times l_c)$ and $U'_{i+1}$ from $[a'_{i+1}, b'_{i+1}]$ to $(a'_{i+1}-p/2\times l_c, b'_{i+1}]$; see the red rectangles in the middle of~\cref{fig:Adamapper-PmSetting}(left) for an illustration. 
\item[C2:] \underline{if  $U'_i\in \Rcal$ and $U'_{i+1}\in \Rcal$},
we expand $U'_i$ from $[a'_i, b'_i]$ to $[a'_i, b'_i+p/2\times l_r)$ and $U'_{i+1}$ from $[a'_{i+1}, b'_{i+1}]$ to $(a'_{i+1}-p/2\times l_r, b'_{i+1}]$; see the blue rectangles in the middle of~\cref{fig:Adamapper-PmSetting}(left) for an example.
\item[C3:] \underline{if  $U'_i\in \Ccal$ and $U'_{i+1}\in \Rcal$ or $U'_i\in \Rcal$ and $U'_{i+1}\in \Ccal$},
we expand $U'_i$ from $[a'_i, b'_i]$ to $[a'_i, b'_i+p\times l_c)$ and $U'_{i+1}$ from $[a'_{i+1}, b'_{i+1}]$ to $(a'_{i+1}-p\times l_c, b'_{i+1}]$; see the green rectangles the middle of~\cref{fig:Adamapper-PmSetting}(left) for an example.
\end{itemize}

The high-level idea behind this adaptive expansion strategy is to extend the initial intervals by a proportion determined by their segment types, namely $\mathcal{C}$ or $\mathcal{R}$. For the C3 case, which lies between C1 and C2, we apply a larger expansion based on $l_c$, using $p$ instead of $p/2$. 
Note that after Step~2, the elements of the cover $\mathcal{U}'$ of $f(X)$ become open intervals.

\para{\underline{Step 3: Finalize the cover $\Ucal$ of $f(X)$}}.
Upon closer inspection of $\Ucal'$, we observe that some intervals $U' \in \Ucal'$ are completely contained within their adjacent intervals. For example, $U'_2$ is contained in $U'_1$, and $U'_6$ is contained in $U'_7$. In this step, we remove such redundant intervals to obtain the final cover $\Ucal$ of $f(X)$, as illustrated at the bottom left of~\cref{fig:Adamapper-PmSetting}. After this refinement, the number of intervals becomes $m=5$.

Using the cover $\Ucal$ of $f(X)$, we then derive the adaptive cover $\Vcal$ of $X$. Based on $\Vcal$, we compute the Mapper graph of $X$ using a user-defined clustering algorithm; see the right panel of~\cref{fig:Adamapper-PmSetting}. Although the choice of clustering algorithm is not central to our method, we additionally provide an automatic parameter selection strategy for DBSCAN based on PD-induced segmentation in Appendix~A.

\subsubsection{Robustness to Imperfect Loop Localization}
\label{sec:impectr}
As noted in~\cref{sec:criticalRange}, the interval $r_i$ derived from the death-triangle strategy may not perfectly localize all points surrounding the feature $\rho_i$. We now show that such imperfect localization is nevertheless sufficient, and in some cases even preferable, for {\adamapper}.
As illustrated in~\cref{fig:criticalRangeInti}, Mapper requires at least three cover elements within a loop-critical region to capture a 1D loop. Given an imperfect interval $r_i=[u_i,v_i]$, the point set
$
\{x \mid f(x)\in [u_i,v_i]\}
$
typically forms a subset of the points surrounding $\rho_i$, ensuring that {\adamapper} allocates at least three intervals within this region.

\begin{figure}[t]
\centering 
\includegraphics[width=\columnwidth]{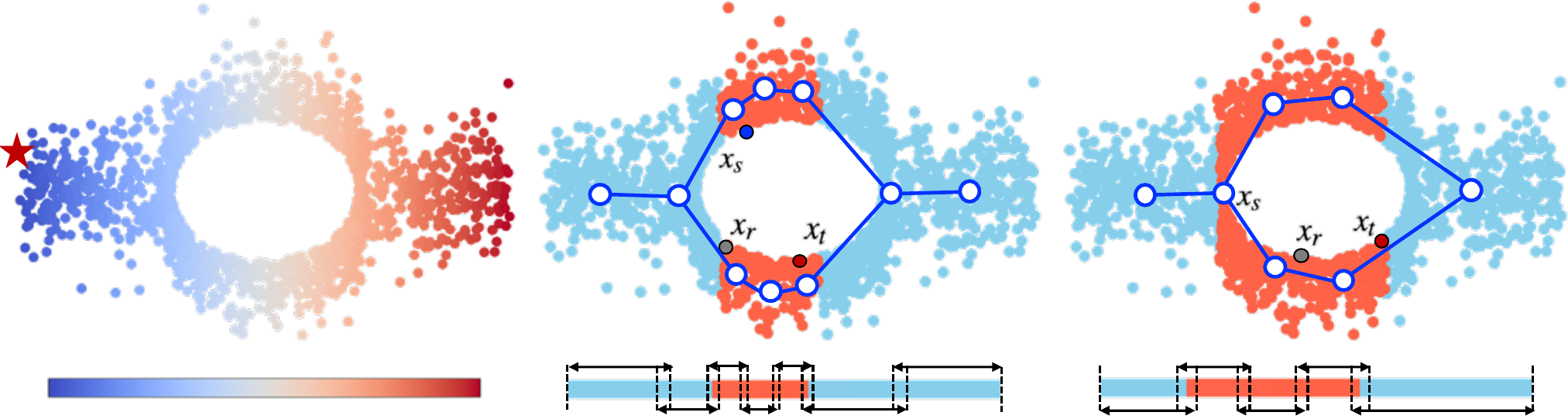}
\vspace{-6mm}
\caption{Capturing loops with an imperfect critical segment. Left: point cloud $X$ colored by the DTB function. Middle and right: segmentations and corresponding Mapper constructions using manually chosen $x_s$ points.}
\label{fig:partial}
\end{figure}

In contrast, cocycle-based strategies~\cite{SilvaMorozovVejdemo-Johansson2011} that include substantially more points may produce excessively large loop-critical regions, particularly when multiple loops interact. When mapped onto the filter function, these enlarged regions can reduce the effective number of intervals assigned to each loop, thereby violating the condition $m \geq 3$ and preventing successful loop capture despite the broader geometric coverage.
Therefore, although the death-triangle strategy may yield only an approximate localization of $\rho_i$, its conservative characterization preserves sufficient resolution in the most relevant regions, enabling {\adamapper} to reliably capture $\rho_i$; see~\cref{fig:partial} for examples.

\subsection{A Toy Example for {\adamapper}}
\label{sec:smpf}

We now consider a slightly more complex example than the point clouds presented previously to illustrate the pipeline of {\adamapper} and demonstrate its ability to capture loops under a user-specified simplification threshold.

\begin{figure}[t]
\centering 
\includegraphics[width=.9\columnwidth]{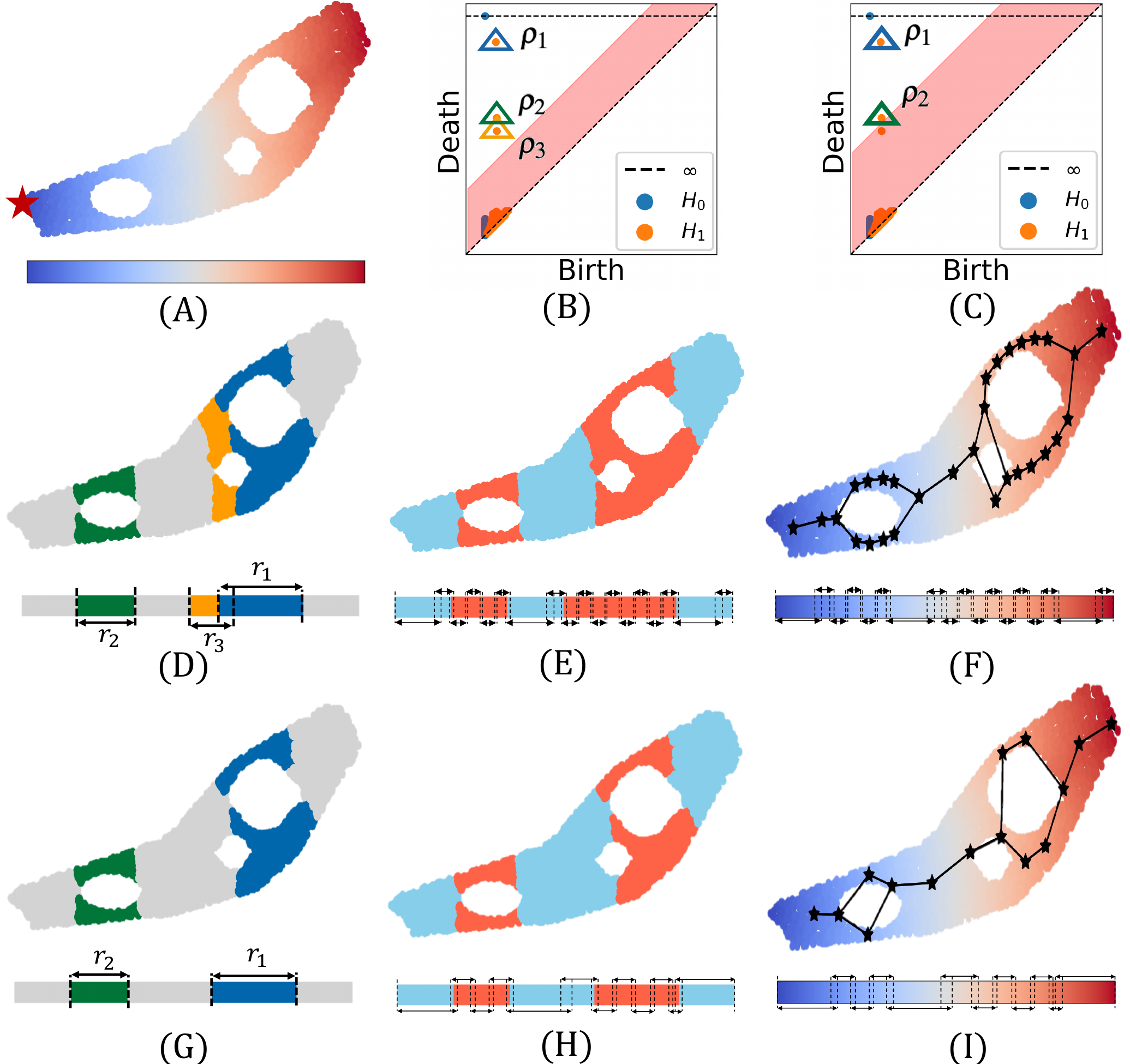}
\vspace{-2mm}
\caption{A toy example illustrating {\adamapper}. (A) Point cloud $X$, colored by the DTB function, with the base point marked by a red star. (B–C) Persistence diagrams with default and user-specified simplification thresholds, $T$ and $T'$. (D–F) and (G–I) Loop-induced ranges, persistence-guided segmentations, and corresponding Mapper constructions for thresholds $T$ and $T'$.}
\vspace{-2mm}
\label{fig:3loops}
\end{figure}

Given the point cloud $X$ in~\cref{fig:3loops}(A), its persistence diagram $\dgm_1(X)$ identifies three loops, $\rho_1$, $\rho_2$, and $\rho_3$, under the default simplification threshold $T$; see~\cref{fig:3loops}(B). Using the DTB function $f(X)$ together with the methodology introduced in~\cref{sec:criticalRange}, we derive the corresponding ranges $r_1$, $r_2$, and $r_3$ along $f(X)$; see~\cref{fig:3loops}(D).
Notably, $r_1$ and $r_3$ overlap. The critical segment $\mathcal{C}$ is therefore defined as
$
r_1 \cup r_2 \cup r_3,
$
while the regular segment $\mathcal{R}$ consists of the remaining portions of $f(X)$; see~\cref{fig:3loops}(E). 
Following the strategy described in~\cref{sec:internalParam}, we set the initial interval length within $\mathcal{C}$ to
$
l_c = |r_3|/3.
$
The resulting cover $\mathcal{U}$ of $f(X)$ is illustrated at the bottom of~\cref{fig:3loops}(E-F). As shown in~\cref{fig:3loops}(F), {\adamapper} successfully captures all three loops using the default parameter setting. 

We now manually choose a simplification threshold $T'$ to filter out the loop $\rho_3$; see~\cref{fig:3loops}(C). Under this threshold, the selected loop ranges reduce to $r_1$ and $r_2$, yielding a critical segment defined by
$
\mathcal{C} = r_1 \cup r_2;
$
see~\cref{fig:3loops}(G-H). 
In this setting, the construction of the cover $\mathcal{U}$ of $f(X)$ begins with the initialization
$
l_c = |r_2|/3,
$
which ultimately produces a Mapper graph containing two loops; see~\cref{fig:3loops}(H-I).

\begin{figure}[t]
\centering 
\includegraphics[width=\columnwidth]{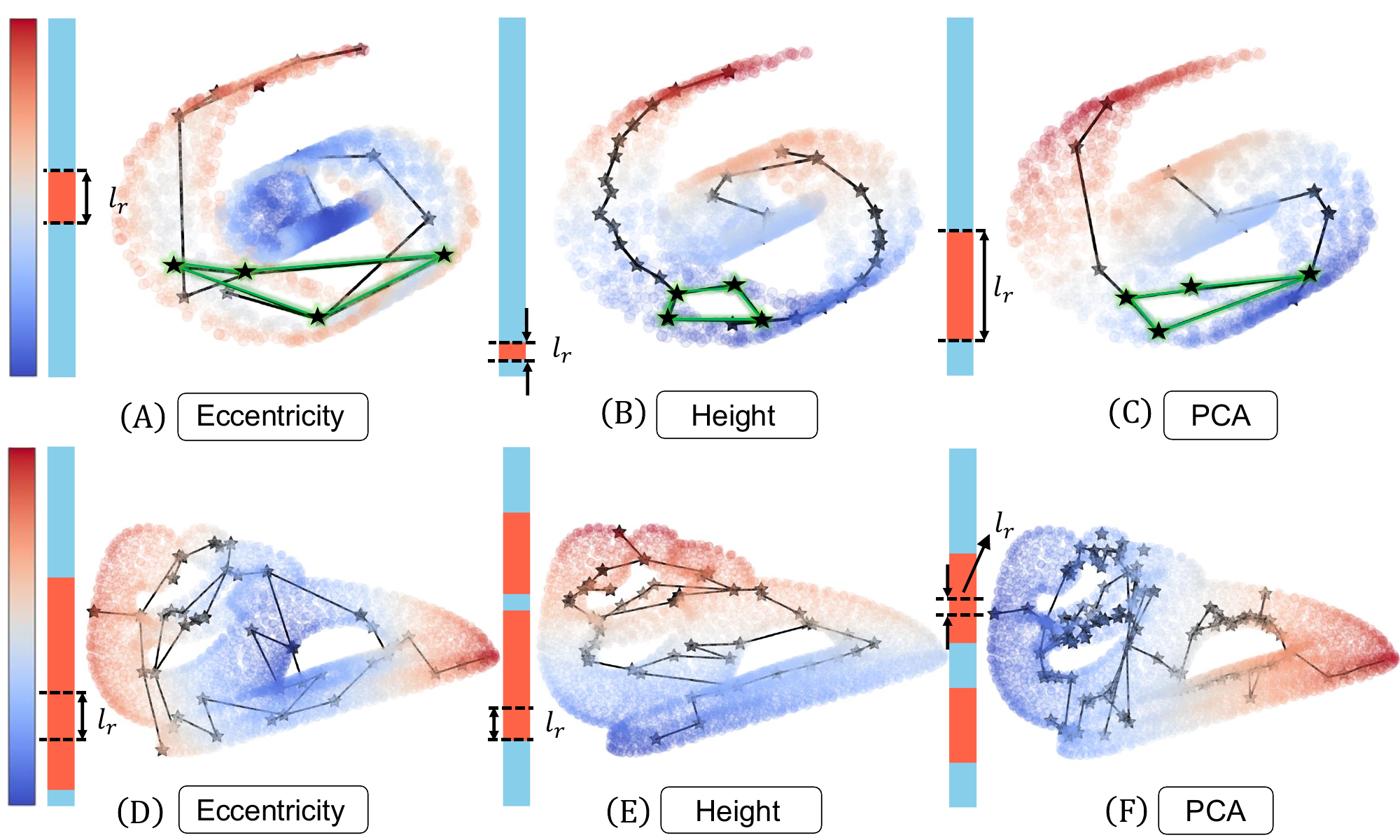}
\caption{Effect of different filter functions $f(X)$ on {\adamapper} constructions.
(A--F) Point clouds colored by filter values (blue: low, red: high); adjacent color bars indicate regular (blue) and critical (orange) ranges. Black graphs show the {\adamapper} skeletons.
(A--C) \SH\ and (D--F) \fertility\ datasets using eccentricity, height, and PCA filters.}
\vspace{-2mm}
\label{fig:3Filterfunction}
\end{figure}

\subsection{Discussion on {\adamapper}}
\label{sec:limAdaMapper}

{\adamapper} is designed to guide Mapper construction using persistent homology as a mechanism for adaptive resolution control, rather than to guarantee recovery of all topological features present in the data. Consequently, its behavior reflects both the strengths and inherent limitations of persistence-guided Mapper constructions.

\para{Scope and limitations.}
Not every loop identified in a persistence diagram can necessarily be represented by {\adamapper}. For example, a torus contains two independent loops, yet the loop associated with the inner tube may fail to appear in the resulting Mapper, even when it is detected through PD-induced segmentation. This limitation stems from the interaction between the filter function and the Mapper cover: some topological features cannot be faithfully represented under a given scalar function, regardless of how finely the cover is refined. Such behavior is a well-known property of Mapper-based summaries and is not specific to {\adamapper}.

\para{Persistence versus Mapper resolution.}
The length of an interval $r_i$ derived from PD-induced segmentation is not necessarily proportional to the persistence of the corresponding feature $\rho_i$. Persistence measures the stability of a feature across scales in the original space, whereas $r_i$ is defined along the filter function. Consequently, persistence-based simplification does not directly correspond to interval-based simplification in Mapper. In particular, the smallest interval $r_i$ may correspond to a feature with relatively high persistence, potentially leading to over-refinement or unintended simplification. To address this discrepancy, {\adamapper} does not assume a one-to-one relationship between persistence and cover size. Instead, it uses persistence to localize loop-critical regions while explicitly controlling resolution along the filter function.

\para{Filter function dependence and sensitivity.}
Like all Mapper-based methods, {\adamapper} depends on the choice of filter function. In this work, we use the distance-to-basepoint (DTB) function as the default, although the framework itself is not restricted to this choice. \Cref{fig:3Filterfunction} presents {\adamapper} constructions using eccentricity, height, and PCA as filter functions. For both the {\SH} and {\fertility} datasets, the same loop-critical features are consistently identified across all filters through PD-induced segmentation.

However, the resulting interval sizes $l_c$ and $l_r$ vary substantially across filters, leading to different cover resolutions and Mapper complexities. This effect is particularly evident in \cref{fig:3Filterfunction}(B) for {\SH} and in \cref{fig:3Filterfunction}(E-F) for {\fertility}, where different filter choices produce noticeably different graph densities, even though all resulting Mapper graphs remain meaningful and interpretable. Importantly, when the filter function varies smoothly along the underlying manifold, {\adamapper} consistently captures the loop-critical regions while making the influence of filter choice on Mapper complexity explicit and transparent.
\section{{\adahisomap}: Adaptive Homology-Preserving Dimensionality Reduction}
\label{sec:AdaHIsomap}

We now introduce {\adahisomap}, an adaptive homology-preserving DR technique that combines {\adamapper}-derived homological skeletons with landmark Isomap. By using homology-informed skeletons to guide landmark selection, {\adahisomap} aims to better preserve both loop structures and global connectivity in low-dimensional embeddings.

\subsection{Adaptive Homology-Preserving Landmark Selection}
\label{sec:landmarking}
{\adahisomap} follows the same overall pipeline as {\hisomap} (\Cref{sec:isomap}), with the key difference occurring in Step 3: landmark construction. Specifically, {\adahisomap} modifies {\hisomap} in two principled ways:
(1) We replace the classical Mapper with {\adamapper}, which constructs a homology-informed skeleton using PD-induced adaptive covers. (2) We augment Mapper-based landmarks with stochastic anchor points sampled from the regular segment $\Rcal$ of the filter function.

Together, these modifications aim to preserve both 1D (loop) and 0D (connectivity) topological features during DR, while substantially reducing the need for manual parameter tuning.

\para{Mapper-based landmarks for 0D and 1D homology preservation.}
In the classic Mapper algorithm, each node represents an abstract cover element of the point cloud, corresponding to a cluster of points in $X$. 
In our framework, following the~{\hisomap} approach, we select the centroid of each cluster of $\Mcal$ (constructed with {\adamapper}) as its representative, and these representatives collectively serve as the subset of the landmarks for~{\adahisomap}. 

\para{Stochastic anchor points for enhancing 0D homology preservation.}
A closer examination of the cover $\Ucal$ of $f(X)$ used in constructing $\Mcal$, such as those shown in~\Cref{fig:3loops,fig:Adamapper-PmSetting}, reveals that substantially more clusters---and therefore more landmarks---are selected within the critical segment $\Ccal$ than within the regular segment $\Rcal$. This imbalance in landmark distribution can distort the preservation of 0D homological features. Importantly, such distortions cannot be resolved simply by decreasing the interval lengths in the regular segment. We illustrate this phenomenon with an example in~\Cref{sec:ToyIsomap}.

To improve the preservation of 0D features, we augment {\adahisomap} with stochastic anchor points. Specifically, we randomly select one data point from each open interval in the regular segment $\Rcal$ and include it as an additional landmark. These stochastic anchors rebalance the landmark distribution and improve the preservation of global connectivity.

Without these additional anchors, most landmarks are concentrated near loop-critical regions, while the regular segment $\Rcal$ remains sparsely represented. Consequently, regions away from dominant loops may be insufficiently sampled during DR, leading to distortions in preserving 0D homological features; see~\Cref{fig:SwissHole}(F). By selecting one stochastic anchor point from each interval in $\Rcal$, we maintain a more balanced representation across the filter range and improve the preservation of global connectivity without introducing a large number of additional landmarks; see~\Cref{fig:SwissHole}(H).

\subsection{A Toy Example}
\label{sec:ToyIsomap}

\begin{figure}[t]
\centering 
\includegraphics[width=\columnwidth]{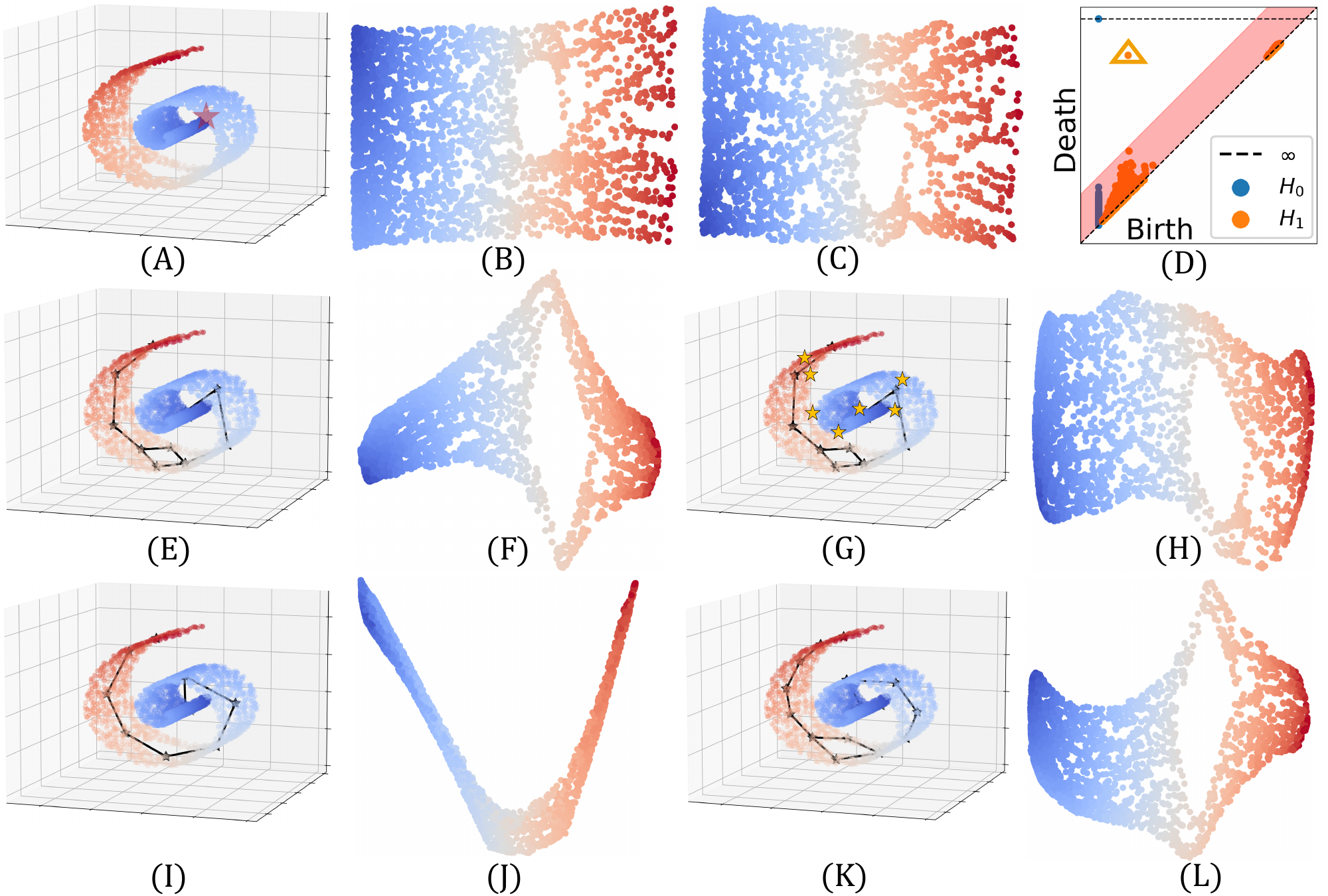}
\vspace{-4mm}
\caption{2D embeddings of the {\SH} dataset. (A) Point cloud $X$, colored by the DTB function $f(X)$. (B) Isomap embedding. (C) Random L-Isomap embedding with $|X_L|=18$. (D) Persistence diagram $\dgm(X)$. (E–F) L-Isomap using nodes of the Mapper computed by {\adamapper} as landmarks. (G–H) {\adahisomap} embeddings with $|X_L|=18$. (I–J) and (K–L) L-Isomap embeddings using nodes of the standard Mapper as landmarks (\hisomap), with $m=10$ and $m=15$.
}
\vspace{-2mm}
\label{fig:SwissHole}
\end{figure}

We illustrate the {\adahisomap} pipeline using the \SH~dataset (\Cref{fig:SwissHole}), a point cloud sampled from a Swiss roll containing a subtle interior loop. For comparison, \Cref{fig:SwissHole}(B-C) show embeddings generated by \isomap~and random \lisomap, respectively, both using $|X_L|=18$ landmarks.

We first compute a DTB filter function with respect to an extremal base point; see~\Cref{fig:SwissHole}(A). We then compute the persistence diagram, which captures the homological structure of the dataset; see~\Cref{fig:SwissHole}(D). Based on the detected loop, {\adamapper} constructs a 1D homological skeleton (the black skeleton in~\Cref{fig:SwissHole}(E)), whose nodes are used as a subset of landmarks for {\adahisomap}.

If only the skeleton nodes are used as landmarks, the resulting 2D embedding is shown in~\Cref{fig:SwissHole}(F). To further improve the preservation of 0D homology, {\adahisomap} introduces stochastic anchor points, highlighted as orange stars in~\Cref{fig:SwissHole}(G). The resulting embedding is shown in~\Cref{fig:SwissHole}(H). Comparing \Cref{fig:SwissHole}(F) and (H) with (B) demonstrates the effectiveness of {\adamapper}-induced landmarks in preserving 1D homology, while also partially preserving 0D homological structure during DR. Furthermore, comparing \Cref{fig:SwissHole}(F) with (H) shows that stochastic anchor points in the regular segment are essential for mitigating distortions of 0D features located away from the dominant loop.

We next investigate why stochastic anchor points are necessary, rather than simply increasing the resolution of {\adamapper} within the regular segment. To this end, we first examine the behavior of the classical Mapper under different parameter settings. Our results show that the number of intervals $m$ must exceed $15$ for the classical Mapper to successfully capture the prominent hole in the \SH~dataset; compare~\Cref{fig:SwissHole}(I) and (K). When fewer intervals are used, the Mapper captures insufficient homological information, leading to greater distortion in the resulting  embedding; compare~\Cref{fig:SwissHole}(J) and (L).

In contrast, {\adahisomap} captures the loop structure using only $m=10$, where the parameter value is determined automatically by {\adamapper} without manual tuning. \Cref{fig:SwissHole}(L) may also be interpreted as the result of a 2D embedding constructed using fine-grained cover elements concentrated within the regular segment. Comparing \Cref{fig:SwissHole}(L) with (H) further demonstrates that stochastic anchor points sampled from the regular segment reduce distortions for data points located away from one-dimensional loops.

Compared with Isomap, {\adahisomap} provides clear advantages in preserving significant 1D loops under simplification, as further demonstrated in~\Cref{sec:results}. Nevertheless, {\adahisomap} retains certain limitations, which are discussed in~\Cref{sec:limAdaHisomap}.

\subsection{Computational Considerations}
Compared with {\hisomap}, {\adahisomap} introduces modest additional computational overhead. The dominant computational costs---including $k$NN graph construction, geodesic distance computation, and multidimensional scaling---remain unchanged from the original Isomap-based pipeline. The additional overhead in {\adahisomap} arises primarily from constructing a homology-informed skeleton, while persistence diagrams are used only to identify prominent 1D features rather than to analyze the full filtration of the dataset. 
In practice, persistence diagram computation can be further accelerated through subsampling, and all experiments in~\Cref{sec:results} were conducted at data scales where this overhead remains manageable. Consequently, {\adahisomap} remains computationally practical for the datasets considered in this work, while providing additional topological guidance for landmark selection.

Beyond algorithmic runtime, practical computational cost is also influenced by parameter-tuning requirements. In~\Cref{sec:results}, baseline methods such as Isomap, UMAP, and t-SNE were evaluated across multiple parameter configurations for each dataset, whereas AdaHIsomap, TopoAE++, and TopoMap were used largely with their default settings (see Appendix~B, Table~II). Therefore, overall computational effort cannot be assessed solely based on runtime comparisons.

\subsection{Limitation Analysis of~{\adahisomap}}
\label{sec:limAdaHisomap}
{\adahisomap} inherits several limitations from Isomap. The most significant is its inability to effectively analyze datasets containing multiple manifolds. For this reason, all datasets considered in~\Cref{sec:results} consist of a single manifold. Nevertheless, compared with Isomap, {\adahisomap} exhibits improved robustness to noise, since noisy points are less likely to be selected as landmarks.

Another limitation stems from the use of classical MDS in Isomap to project the geodesic distance matrix. This projection is fundamentally a linear transformation based on pairwise distances. Consequently, when loops encoded in the distance matrix are oriented approximately perpendicular to one another, some loops may not be preserved in the resulting Isomap embedding. The same limitation naturally carries over to {\adahisomap}; see, for example,~\Cref{fig:Fertility}.

In addition, the performance of {\adahisomap} depends strongly on the quality of the {\adamapper} construction, making it sensitive to limitations inherited from that component. Unlike {\adamapper}, {\adahisomap} may not fully reflect the intended simplification effect, since landmarks can still be placed near low-persistence loops.

Finally, {\adahisomap} is intended as a homology-informed extension of landmark Isomap rather than a comprehensive solution to all topology-preserving dimensionality reduction problems. Future work could incorporate more advanced embedding techniques or alternative distance models to further address these limitations.

\section{Experimental Results} 
\label{sec:results}

We first describe the experimental setup, baseline methods, and parameter settings used to evaluate our approach. 

\begin{figure}[t]
\centering 
\includegraphics[width=\columnwidth]{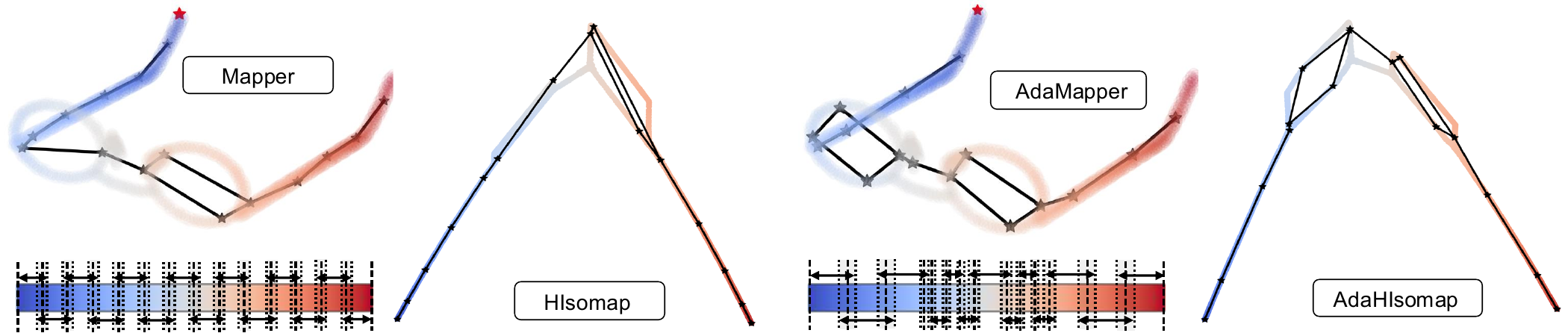}
\vspace{-6mm}
\caption{Comparison of fixed-cover and adaptive-cover Mapper constructions and their impact on downstream dimensionality reduction.}
\vspace{-2mm}
\label{fig:hisomapVS}
\end{figure}

\para{Evaluation rationale.}
Before presenting the dataset-specific results, we first clarify the evaluation rationale underlying our experimental design. As illustrated in~\Cref{fig:hisomapVS}, fixed-cover Mapper constructions---and consequently {\hisomap}---are highly sensitive to manually specified cover resolutions. Even reasonable parameter choices can produce substantially different Mapper skeletons and downstream embeddings, making direct comparisons based on ``best'' parameters inherently ill-posed. 
Importantly, \Cref{fig:hisomapVS} is not intended as a performance comparison, but rather as a conceptual demonstration of this parameter sensitivity and its implications for evaluation. In contrast, {\adamapper} automatically derives cover resolutions from persistent homology, yielding a more stable and reproducible structural summary. This observation motivates our evaluation protocol and explains why we focus on comparing {\adahisomap} with methods that do not rely on manual Mapper parameter tuning.

\para{Experimental setup.}
We evaluate our method on 11 datasets to demonstrate its behavior under default settings across diverse data types. All results were obtained on a MacBook Pro with 16 GB of RAM, an Apple M3 chip with 8 CPU cores and 10 GPU cores.
For all datasets except the \coauthor~dataset, we use the default settings of {\adahisomap}. 
Since the \coauthor~dataset contains only link information, we adapt our framework accordingly.
For comparison, we include widely used DR methods introduced in \Cref{sec:related}, selecting the best-performing configurations through parameter sampling. 
For TopoAE++, we use the default settings reported in~\cite{ClemotDigneTierny2025}. 
TopoMap, which is primarily designed to preserve 0D homological features, performs poorly relative to the other methods. Its embeddings are therefore presented in Appendix C, except in cases where TopoAE++ results are unavailable (\bcsstk~due to its size and \coauthor~due to its data type).

We omit explicit \hisomap~results. Since \hisomap~requires manual parameter specification, running it with the cover parameters automatically derived by \adamapper~produces nearly identical outcomes, while alternative parameter choices can yield substantially different results.

A full list of parameter settings is given in Appendix B, Table II. In the remainder of this section, we use N to denote dataset size and Dim to denote dimensionality.

\para{Quantitative evaluation.}
Table~\ref{tab:EVAtable} summarizes the quantitative evaluation results across all datasets, with evaluation-stage normalization applied to all datasets except \coauthor, as described in~\Cref{sec:Eva_Metrics}. For each metric, the best result is highlighted in \textbf{bold}, and the second-best result is \underline{underlined}. The evaluation is based on three metrics: one measuring geometric fidelity (RMSE) and two measuring topological preservation (PDW$^0$ and PDW$^1$). Detailed definitions of these metrics are provided in~\Cref{sec:Eva_Metrics}.

\begin{table}[!t]
\captionsetup{justification=raggedright, singlelinecheck=false} 
\scriptsize  
\raggedright 
\setlength{\tabcolsep}{3pt}  
\renewcommand{\arraystretch}{0.9}  
\caption{Quantitative evaluation results for 11 datasets across five methods, with evaluation-stage normalization applied to all datasets except \coauthor. WD1 denotes ${PDW}^{1}(X, Y)$, and WD0 denotes ${PDW}^{0}(X, Y)$.}
\resizebox{\columnwidth}{!}{%
\begin{tabular}{l l cccccccc}
\toprule

\textbf{Dataset} & \textbf{Metric} & \textbf{\adahisomap} & \textbf{\isomap} & \textbf{T-SNE} & \textbf{UMAP} & \textbf{\topomap} & \textbf{TopoAE++} \\
\midrule

\multirow{4}{*}{\glasses~\cite{LopezFuertesCabrera2013,CabreraSastreRodriguez2012,LiGodilTatsumaYanagimachi2012,ShilaneMinKazhdanFunkhouser2004,BronsteinBronsteinKimmel2008, SumnerPopovic2004}}
& RMSE & \textbf{4.4721} & 10.0819 &  \underline{5.5015}  &  5.5312 & 9.5470 & 5.8106\\
& WD1  & 2.3330 & \textbf{2.2930} &   2.6935  &  \underline{2.3318} & 2.6638 & 2.6500\\
& WD0  & 15.4884 & 16.1414 & 14.2706 & 15.5181 & \textbf{13.1878} & \underline{13.5380}\\
\midrule

\multirow{4}{*}{\fertility~\cite{Falcidieno2007}}
& RMSE & 5.4661 & \textbf{2.9248} & 5.1855  &   \underline{3.1769} & 9.4070 & 7.8328\\
& WD1  & \underline{6.8148} & 7.0455 &   7.9490  &  7.2081 & 7.3557 & \textbf{6.7560}\\
& WD0  & 30.7790 & \underline{30.4625} & 31.4957 & \textbf{28.7413} & 44.7255 & 35.4177\\
\midrule

\multirow{4}{*}{\Octa}
& RMSE & 5.3117 & 6.2988 &  \underline{5.1494}  &  \textbf{4.4168} & 13.3433 & 11.9366\\
& WD1  & \underline{8.8498} & 9.1891 &   9.7950  &  9.3686 & \textbf{8.5730} & 9.1632\\
& WD0  & 45.9230 & 45.7363 & \underline{39.4204} & \textbf{39.3655} & 53.7894 & 47.6841\\
\midrule

\multirow{4}{*}{\VS~\cite{Popinet2004,GuentherGrossTheisel2017}}
& RMSE & \textbf{135.2299} & 138.3632 &  \underline{136.0178}  &  136.3662 & 137.2030 & 137.2617\\
& WD1  & 15.5263 & \underline{15.4714} &   \textbf{15.4537}  &  16.0610 & 15.9942 & 15.9470\\
& WD0  & \underline{255.3856} & \textbf{253.9320} & 255.5551 & 258.8871 & 256.0890 & 258.4001\\
\midrule

\multirow{4}{*}{\Mice~\cite{SmarrZuckerKriegsfeld2016}}
& RMSE & \textbf{58.1075} & 59.7248 &  \underline{58.3452}  &  58.5394 & 59.4077 & 60.6419\\
& WD1  & 36.1554 & \underline{35.9450} &   36.2990  &  36.0612 & \textbf{35.5404} & 36.2929\\
& WD0  & 407.4759 & \underline{405.7253} & 409.0985 & 408.1793 & \textbf{404.3524} & 410.8558\\
\midrule

\multirow{4}{*}{\fourelt~\cite{GansnerKorenNorth2005}}
& RMSE & 6.7877 & \textbf{4.3194} &  7.5606  &  \underline{5.9708} & 24.4609 & 9.5671\\
& WD1  & \textbf{8.2003} & 10.1485 &   10.4036  &  9.8665 & \underline{8.2998} & 8.9554\\
& WD0  & 55.3615 & \underline{52.3128} & 54.9673 & \textbf{42.5516} & 68.9178 & 53.5470\\
\midrule

\multirow{4}{*}{\bcsstk~\cite{DuffGrimesLewis1989}}
& RMSE & 6.3713 & \textbf{3.8359} &  5.9974  &  \underline{4.9440} & 12.0154 &  - \\
& WD1  & 8.8243 & 8.7969 &   \textbf{8.1602}  &  9.6296 & \underline{8.4452} &  - \\
& WD0  & \underline{50.0520} & 50.1318 & 51.5510 & \textbf{50.0258} & 61.1355 & - \\
\midrule

\multirow{4}{*}{\GIF~\cite{GIF}}
& RMSE & \underline{23.5719} & 23.9283 &  \textbf{23.3440}  &  23.7147 & 23.6332 & 23.7677\\
& WD1  & 2.3970 & 2.3273 &   2.3658  &  \textbf{2.3163} & \underline{2.3263} & 2.3617\\
& WD0  & \underline{76.0661} & \textbf{75.7054} & 76.2669 & 76.6828 & 76.7583 & 76.6847\\
\midrule

\multirow{4}{*}{\face~\cite{Face3D}}
& RMSE & \textbf{177.9032} & 179.4126 &  178.7576  &  178.8818 & \underline{178.4508} & 179.2001\\
& WD1  & 8.3507 & 8.2292 &   \textbf{8.0623}  &  \underline{8.1025} & 8.1389 & 8.3904\\
& WD0  & \underline{246.0745} & \textbf{245.6767} & 246.1492 & 246.9120 & 248.0108 & 247.0201\\
\midrule

\multirow{4}{*}{\coauthor~\cite{Newman2006}}
& RMSE & - & - &  -  &  - & - & -\\
& WD1  & 13.1683 & \underline{12.8986} &   14.4222  &  \textbf{11.9474} & 13.7099 & -\\
& WD0  & 108.1479 & 94.6187 & 117.7190 & \underline{65.2374} & \textbf{0.0116} & -\\
\midrule

\bottomrule
\end{tabular}
}
\label{tab:EVAtable}
\end{table}

\begin{figure}[t]
\centering 
\includegraphics[width=\columnwidth]{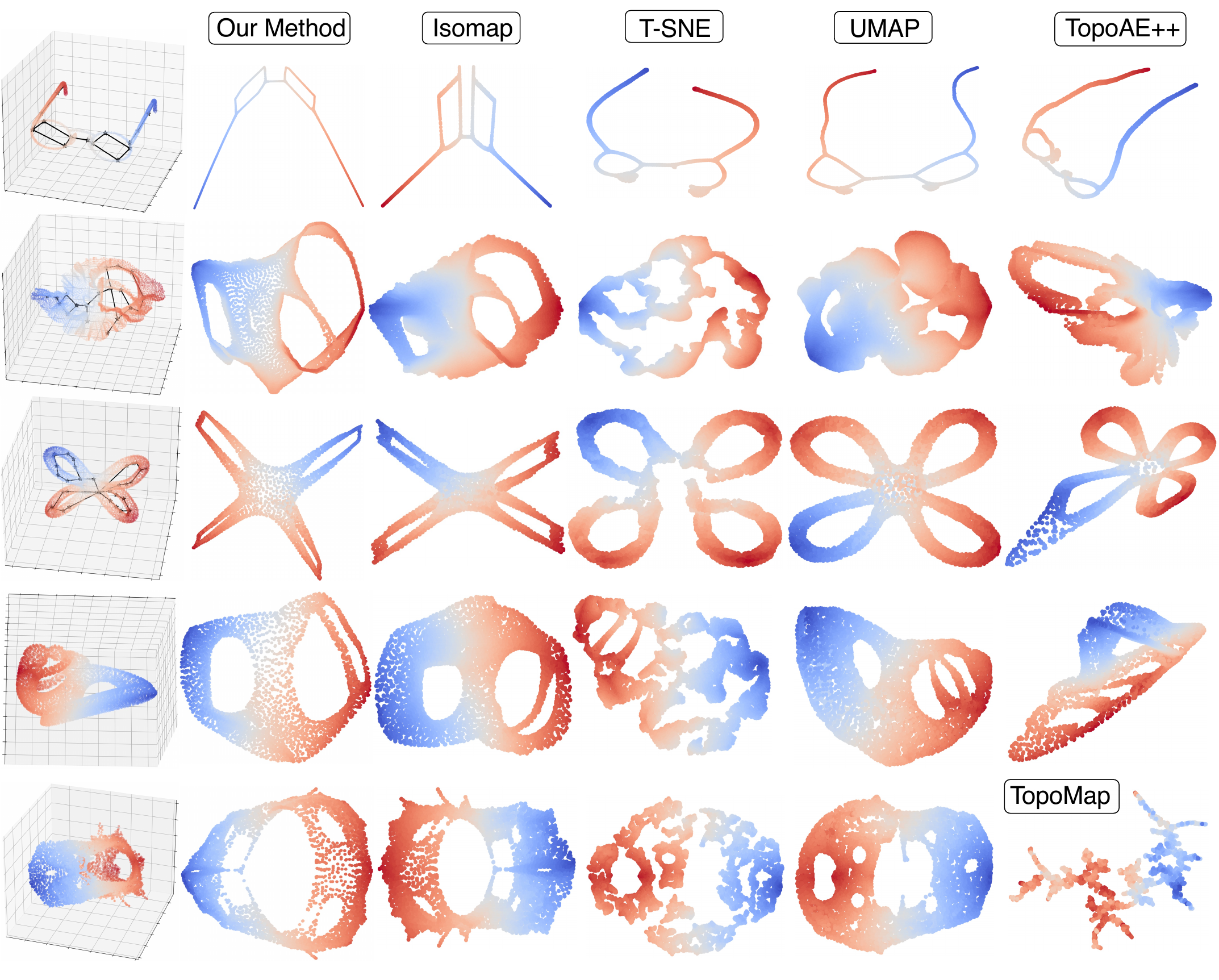}
\vspace{-6mm}
\caption{Dimensionality reduction results for the \glasses, \fourelt, \Octa, \fertility, and \bcsstk\ datasets. From left to right: original point clouds, {\adahisomap}, Isomap, t-SNE, UMAP, and TopoAE++/TopoMap embeddings. Points are colored by the filter function.} 
\vspace{-2mm}
\label{fig:3DShape_4elt_octa_glasses}
\end{figure}

\subsection{Point Cloud Data}
We evaluate our method on point cloud datasets with nontrivial homology, covering 3D and high-dimensional cases. 

\begin{figure}[t]
\centering 
\includegraphics[width=\columnwidth]{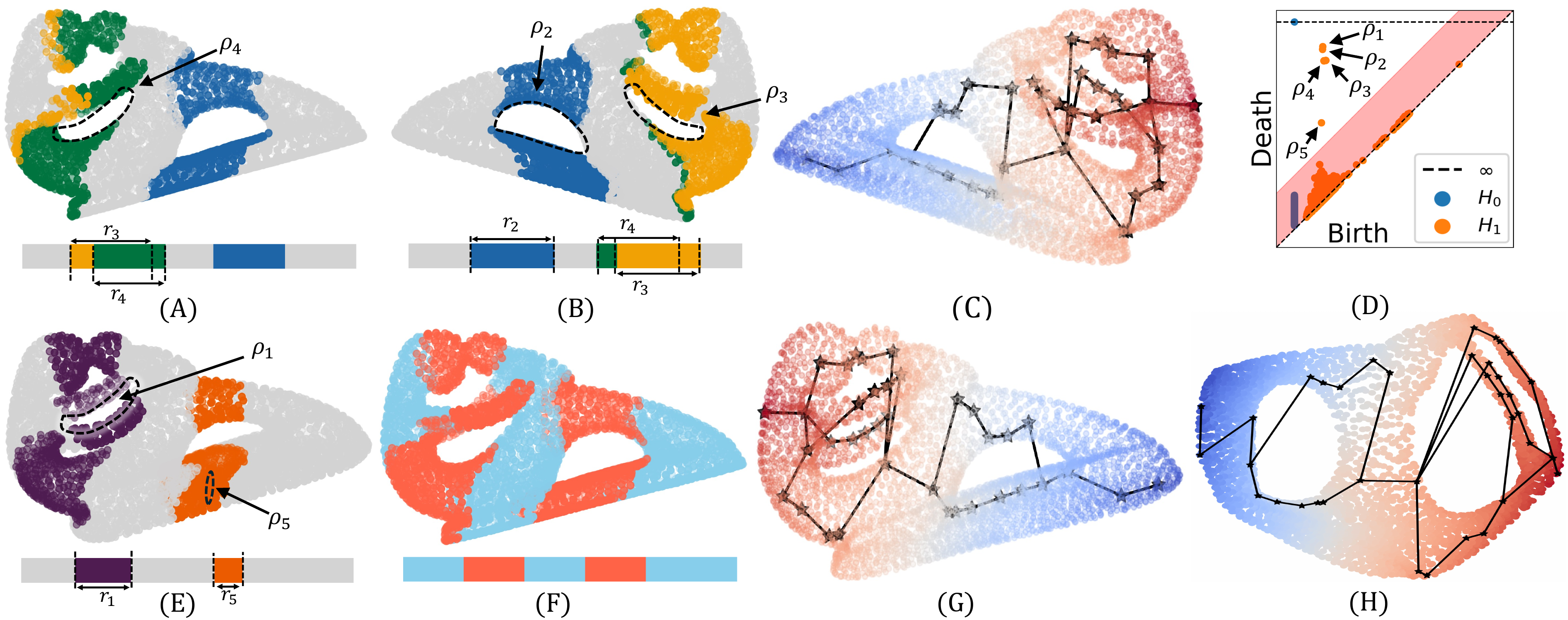}
\vspace{-6mm}
\caption{Mapper analysis of the {\fertility} dataset. (A–B) Three critical ranges captured by {\adamapper} and preserved by {\adahisomap}. (E) Two critical ranges captured by {\adamapper} but not preserved by {\adahisomap}. (D) Persistence diagram. (F) Persistence-guided segmentation. (C, G) Mapper embedded in 3D. (H) Mapper in the {\adahisomap} embedding.} 
\label{fig:Fertility}
\end{figure}

\begin{figure}[t]
\centering 
\includegraphics[width=0.9\columnwidth]{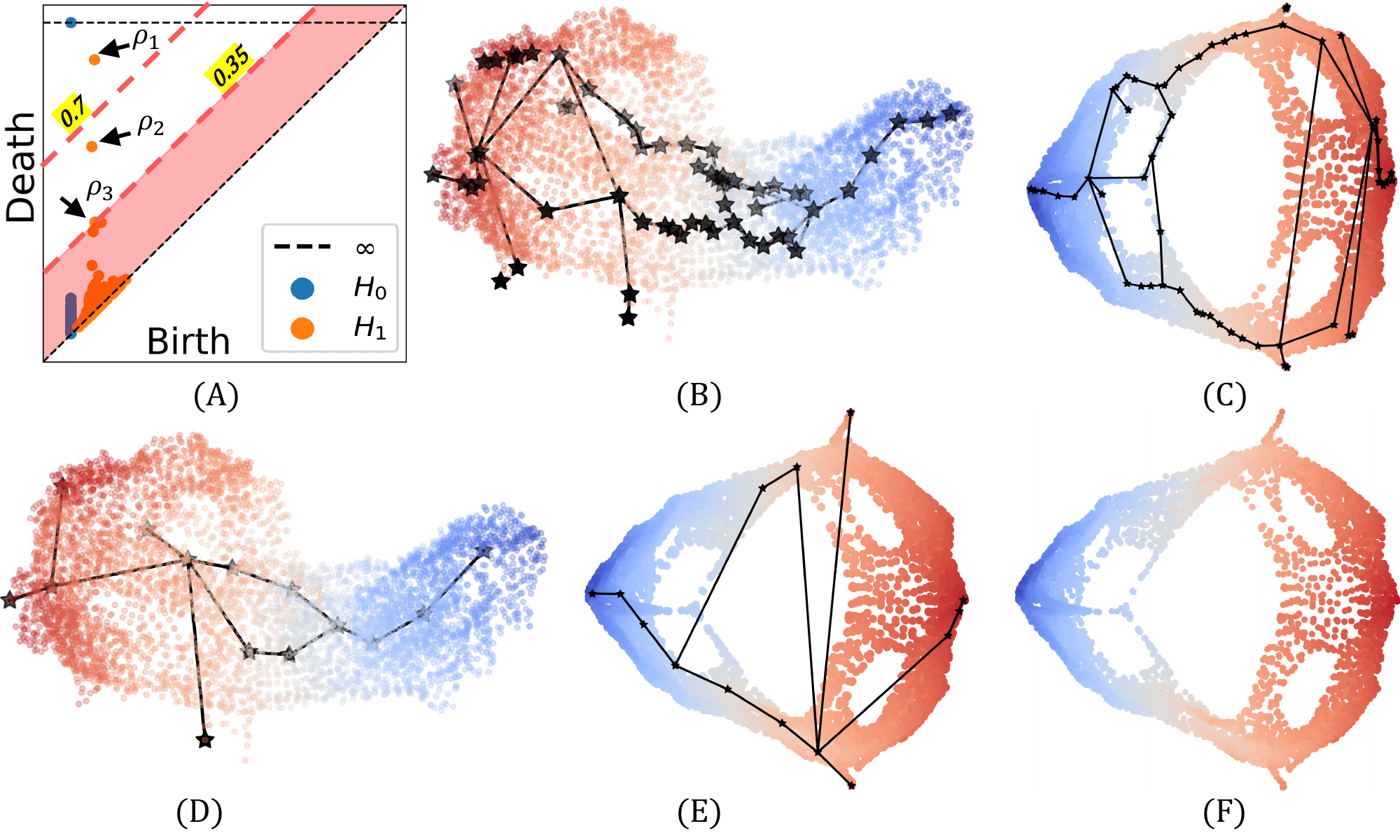}
\vspace{-2mm}
\caption{Mapper analysis of the {\bcsstk} dataset with default and user-specified simplification thresholds $T=0.35$ and $T'=0.7$. (A) Persistence diagram with thresholds $T$ and $T'$. (B–C) Mapper with $T$ in 3D and 2D. (D–E) Mapper with $T'$ in 3D and 2D. (F) {\adahisomap} embedding with $T'$. All visualizations (B–F) are colored by the DTB function.} 
\label{fig:bcsstk}
\end{figure}

\begin{figure}[t]
\centering 
\includegraphics[width=0.9\columnwidth]{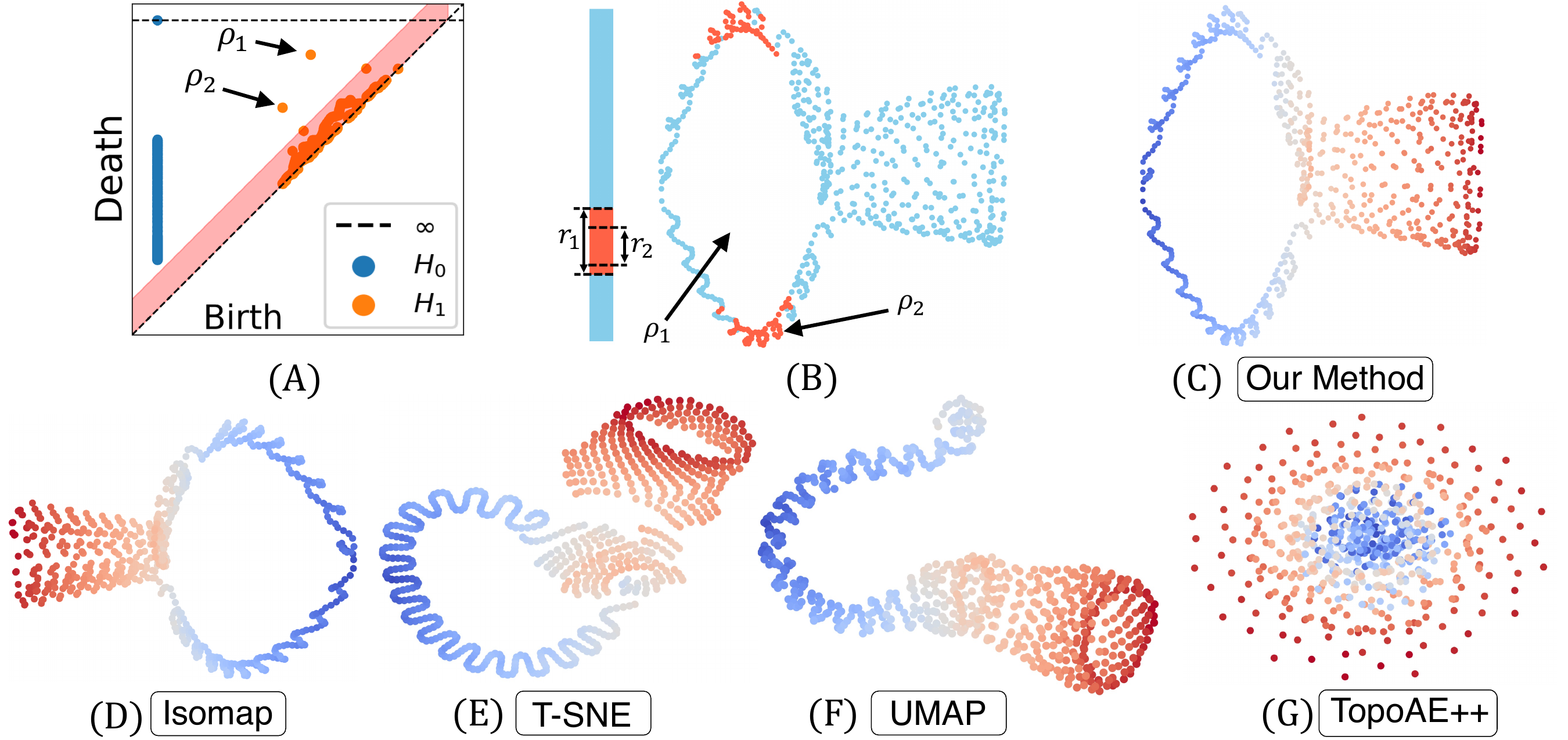}
\caption{Dimensionality reduction of the {\Mice} dataset. (A) Persistence diagram. (B) Persistence-guided segmentation. (C–G) Embeddings produced by {\adahisomap}, Isomap, t-SNE, UMAP, and TopoAE++, respectively, colored by the DTB function.} 
\vspace{-2mm}
\label{fig:Mice}
\end{figure}

\subsubsection{3D Datasets}
 The \mm{\mathsf{Glasses}} dataset is a 3D object with $N=2000$ points~\cite{LopezFuertesCabrera2013}. As shown in \Cref{fig:3DShape_4elt_octa_glasses}(1st row), {\adahisomap} and TopoAE++ outperform the other methods. The remaining methods capture two loops but distort the shape: Isomap and UMAP introduce slight distortion, while t-SNE produces more severe distortion.

The \mm{\mathsf{4elt}} dataset is a 3D embedding with $N=7807$ points~\cite{GansnerKorenNorth2005}. In~\Cref{fig:3DShape_4elt_octa_glasses}(2nd row), both {\adahisomap} and Isomap preserve three major loops and outperform t-SNE, UMAP, and TopoAE++. Quantitatively, {\adahisomap} achieves the best {PDW}$^1$ score, while Isomap and t-SNE perform better on {PDW}$^0$ (see Table~\ref{tab:EVAtable}).  

The \mm{\mathsf{Octa}} dataset has $N=2994$ points sampled from a mesh of octahedral handles. As shown in~\Cref{fig:3DShape_4elt_octa_glasses}(3rd row), {\adahisomap} successfully preserves all four loops. Other methods also perform well, except t-SNE, which struggles with central 0D features.  

The \mm{\mathsf{Fertility}} dataset is a 3D object consisting of $N=3739$ points~\cite{Falcidieno2007}. We first analyze this dataset using \adamapper. As shown in~\Cref{fig:Fertility}(D), the persistence diagram reveals five significant loops.
Specifically, $\rho_1$ is formed by the two arms (\Cref{fig:Fertility}(E)); $\rho_2$ corresponds to the loop between the mother’s legs and the ground (\Cref{fig:Fertility}(B)); $\rho_3$ is enclosed by the left arm and leg (\Cref{fig:Fertility}(B)); $\rho_4$ is enclosed by the right arm and leg (\Cref{fig:Fertility}(A)); and $\rho_5$ corresponds to a tunnel within the 3D object (\Cref{fig:Fertility}(E)).
These results demonstrate that {\adamapper} successfully derives the corresponding ranges for all five significant 1D features, leading to an accurate critical segmentation for these loops; see~\Cref{fig:Fertility}(F). The primary limitations are the inability of both {\adamapper} and {\adahisomap} to capture $\rho_5$, as well as the failure of {\adahisomap} to preserve $\rho_1$.

As shown in \Cref{fig:3DShape_4elt_octa_glasses}(4th row), UMAP performs best, preserving four loops, while TopoAE++ also preserves four loops but with greater distortion of the natural shape.
Although the \adamapper~skeleton captures four loops (\Cref{fig:Fertility}(H)), {\adahisomap} preserves only three in its embedding.
Both UMAP and TopoAE++ outperform {\adahisomap} in preserving $\rho_1$. Because $\rho_1$ is perpendicular to $\rho_3$ and $\rho_4$, {\adahisomap} prioritizes preserving $\rho_3$ and $\rho_4$ at the expense of $\rho_1$, reflecting the limitations of {\adahisomap} discussed in~\Cref{sec:limAdaHisomap}. 
However, based on the evaluation results in Table~\ref{tab:EVAtable}, TopoAE++ and~{\adahisomap} outperforms all other methods on PDW$^{1}$, while UMAP achieves the best performance on PDW$^{0}$. It is worth noting that all DR methods fail to preserve $\rho_5$, which corresponds to a tunnel.

The \mm{\mathsf{Bcsstk31}} dataset represents a 3D embedding constructed from a stiffness matrix with $N=5660$ points~\cite{DuffGrimesLewis1989}.
As shown in~\Cref{fig:3DShape_4elt_octa_glasses}(5th row), Isomap and {\adahisomap} perform better at preserving 1D homological features compared to t-SNE, UMAP, and TopoMap. In particular, {\adahisomap} emphasizes 1D features by enlarging loops, though at the cost of introducing distortions to 0D features.
As illustrated in \Cref{fig:bcsstk}(A), the dataset contains three loops above the default simplification threshold. 
By setting a manual threshold of $T' = 0.7$, two 1D features are filtered out. 
The Mapper generated by \adamapper~is successfully simplified under this homology-based resolution control (\Cref{fig:bcsstk}(D)). 
Applying {\adahisomap} to the simplified Mapper yields the embeddings shown in \Cref{fig:bcsstk}(E) and (F). 
Compared with \Cref{fig:bcsstk}(C), the {\adahisomap} embedding with threshold $T'$ preserves fewer 1D features.
However, the effect of simplification is less pronounced for {\adahisomap} than for \adamapper, since more than two loops remain visible in \Cref{fig:bcsstk}(F). This occurs because several landmarks are located near low-persistence loops, which promote the preservation of local structures around nearby points.

\subsubsection{High-Dimensional Dataset}
The \mm{\mathsf{Mice}} dataset contains $N=674$ points in $Dim=300$~\cite{SmarrZuckerKriegsfeld2016}. In \Cref{fig:Mice}, we compare the DR results together with the persistence diagram computed from the original high-dimensional space. As shown in \Cref{fig:Mice}(A), the dataset contains two significant topological features, $\rho_1$ and $\rho_2$. These features likely correspond to periodic patterns in the mice temperature profiles, with $\rho_1$ associated with circadian rhythms and $\rho_2$ associated with ultradian rhythms.
As shown in \Cref{fig:Mice}(C-G), \adahisomap, Isomap, and UMAP each preserve only one of the two features, whereas t-SNE successfully preserves both and achieves the best overall performance. TopoAE++ fails on this dataset.
Interestingly, \Cref{fig:Mice}(B) shows that {\adahisomap} attempts to preserve $\rho_2$ by incorporating $|r_2|$ into the Mapper construction, since $|r_2| < |r_1|$. However, as discussed in~\Cref{sec:limAdaMapper}, \adamapper~cannot guarantee the preservation of all features detected in the persistence diagram. Furthermore, \Cref{fig:Mice}(B) reveals that the critical range $r_1$ associated with $\rho_1$ in the DTB function is imperfect. Nevertheless, as discussed in~\Cref{sec:impectr}, even an imperfect PD-induced segmentation can still yield a meaningful Mapper construction.

\subsection{Scientific Data}
\begin{figure}[t]
\centering 
\includegraphics[width=0.9\columnwidth]{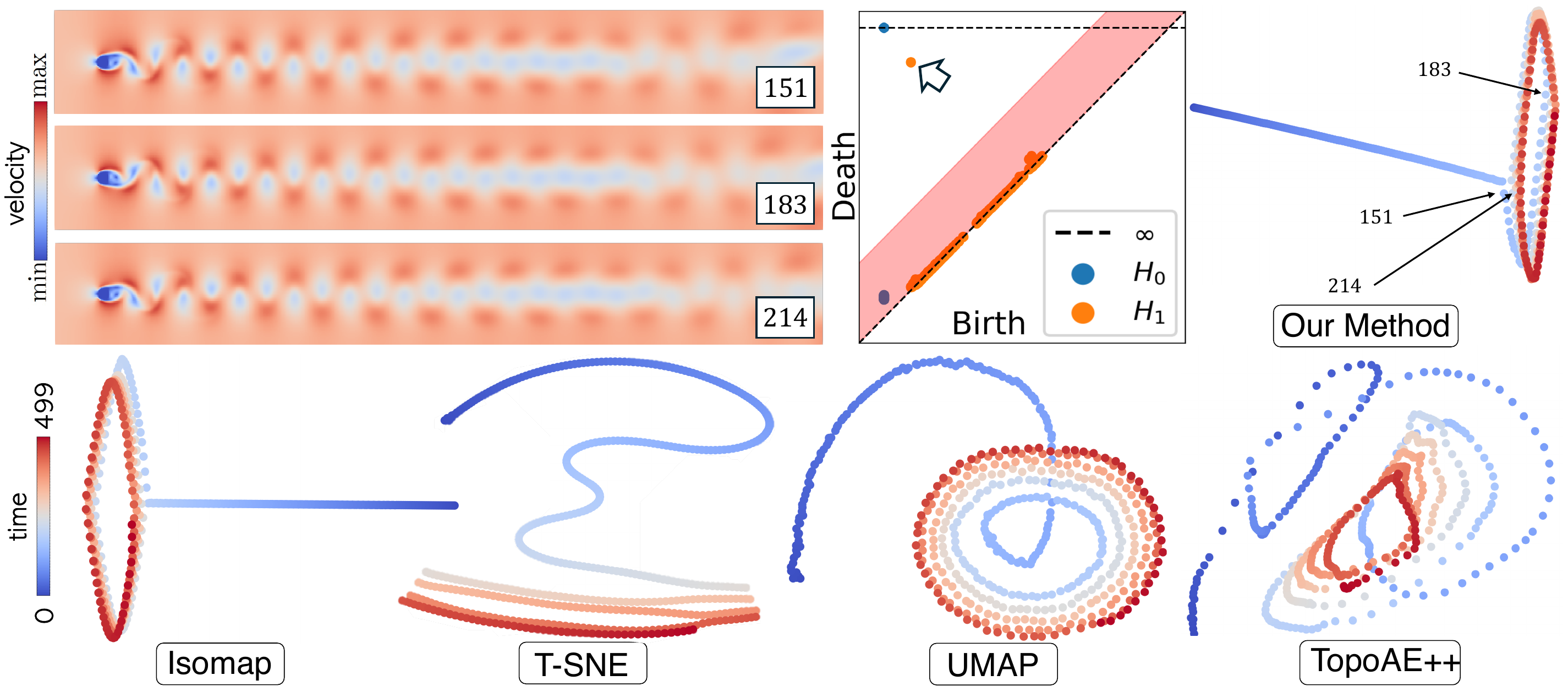}
\caption{Dimensionality reduction for the {\VS} dataset. Top left: selected instances. Top middle and right: persistence diagrams in the original space and the {\adahisomap} embedding, respectively. Bottom: embeddings produced by Isomap, t-SNE, UMAP, and TopoAE++.} 
\vspace{-2mm}
\label{fig:VS}
\end{figure}

The \mm{\mathsf{Vortex Street}} dataset is a classic 2D von K\'arman vortex street simulation~\cite{Popinet2004,GuentherGrossTheisel2017} with $N=495$ and $Dim=8192$, where each simulation is treated as a data point.
We focus on the region of vortex shedding behind the cylinder and use velocity magnitude as the scalar field. Since vortex shedding becomes intense after 1005 time steps, we analyze the last 495 time instances (steps 1005–1500 of the original 1500). 
This subset emphasizes the periodic flow structures induced by vortex shedding and serves as an effective test case for evaluating the ability of DRs to capture recurring topological features.
The persistence diagram (\Cref{fig:VS}, top middle) reveals a prominent 1D homological feature.

The 2D embeddings of four DR algorithms are shown in~\Cref{fig:VS}(bottom), where each point represents an instance and is colored by time.
All DR algorithms detect the periodicity from steps 151 to 495, although TopoAE++ exhibits greater distortion of the natural shape.
The results indicate a repeating pattern of alternating vortices with a time interval roughly equal to 63. 
For example, the instance at time step 151 is very similar to the instance 214. 
Among the embeddings, Isomap and {\adahisomap} provide the clearest separation between periodic and non-periodic regions.
Furthermore, as shown in Table~\ref{tab:EVAtable}, {\adahisomap} achieves the best RMSE compared to the other methods.

\subsection{Imaging Data}
\begin{figure}[t]
\centering 
\includegraphics[width=\columnwidth]{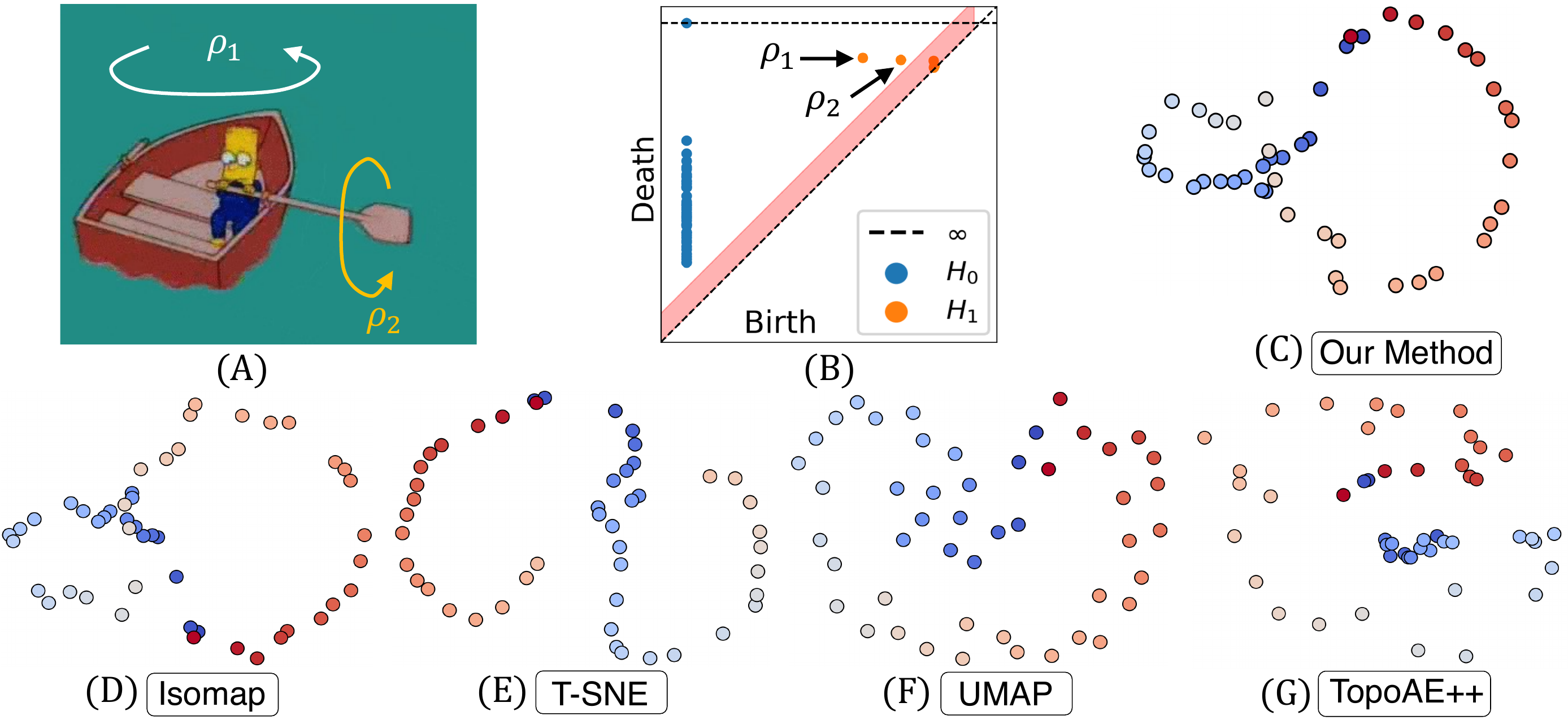}
\caption{Dimensionality reduction of the {\GIF} dataset. (A) Example instance. (B) Persistence diagram. (C–G) Embeddings produced by {\adahisomap}, Isomap, t-SNE, UMAP, and TopoAE++, respectively, all colored by time.} 
\vspace{-2mm}
\label{fig:cartoon}
\end{figure}

The \mm{\mathsf{Cartoon}} dataset records a cartoon character rowing a boat and consists of $N=48$ frames with $Dim=4900$~\cite{GIF}. This dataset is of interest because it contains two nontrivial 1D loops, $\rho_1$ and $\rho_2$ (\Cref{fig:cartoon}(B)). The dominant feature, $\rho_1$, corresponds to the circular motion of the character with the boat, while $\rho_2$ arises from the circular motion of the paddle (\Cref{fig:cartoon}(A)). As shown in \Cref{fig:cartoon}(C--G), most DR methods are able to preserve both $\rho_1$ and $\rho_2$.
However, only {\adahisomap} preserves these two features using its default parameter settings. For the other methods, we explored a wide range of parameter combinations to obtain embeddings that reveal both loop structures.

\begin{figure}[t]
\centering 
\includegraphics[width=\columnwidth]{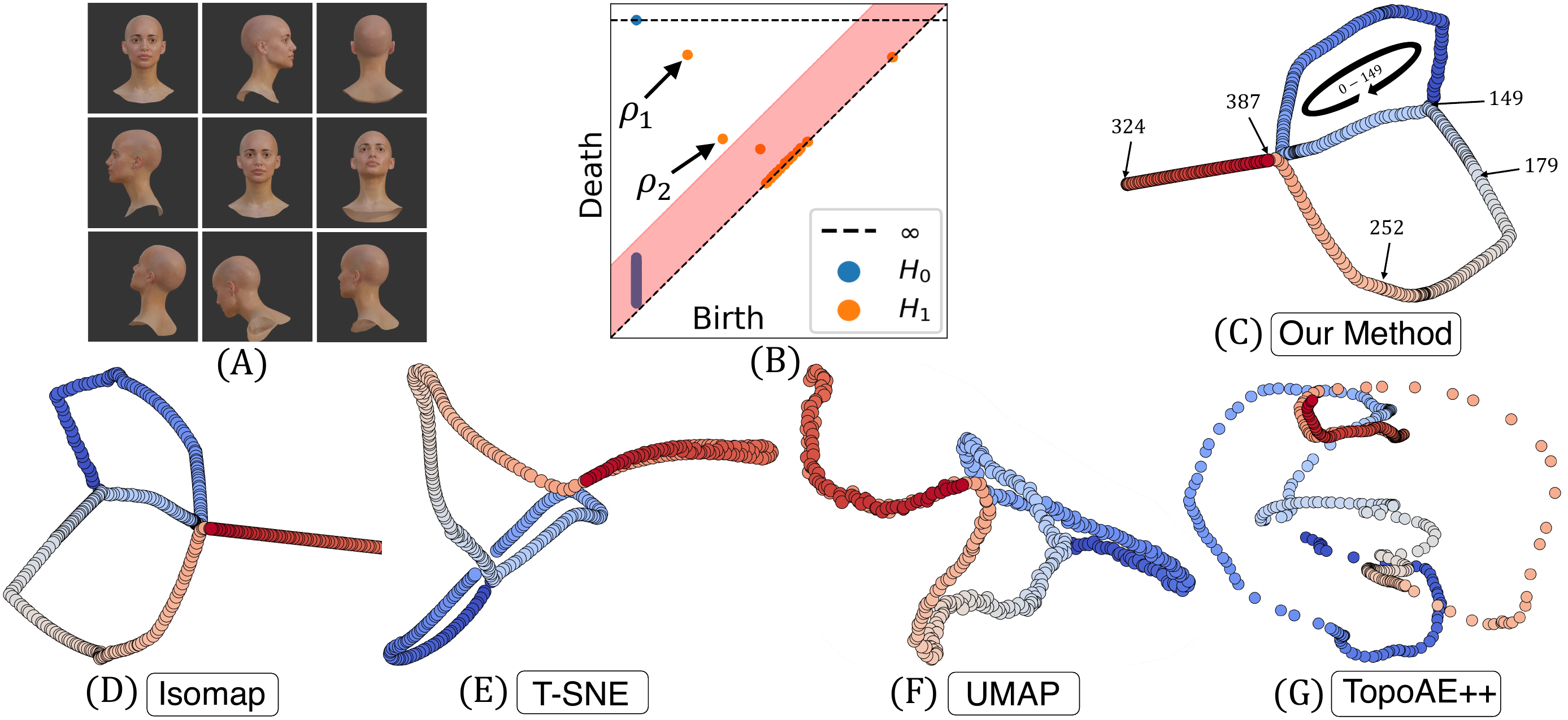}
\caption{Dimensionality reduction of the {\face} dataset. (A) Nine sample instances. (B) Persistence diagram. (C–G) Embeddings produced by {\adahisomap}, Isomap, t-SNE, UMAP, and TopoAE++, respectively, all colored by time.} 
\label{fig:face}
\end{figure}

The \mm{\mathsf{Face 3D Model}} dataset is constructed by capturing a face from different viewing angles and consists of $N=387$ frames with $Dim=4900$~\cite{Face3D}. The camera first moves horizontally around the subject from time steps 0–149, then moves vertically downward from 149–179, shifts to the right from 179–252, moves upward from 252–324, and finally moves downward again from 324–387.
As shown in \Cref{fig:face}, the persistence diagram identifies two significant loops, $\rho_1$ and $\rho_2$. Both {\adahisomap} and Isomap perform well in preserving $\rho_1$, which corresponds to the horizontal rotation from time steps 0–149. The second feature, $\rho_2$, does not represent a full rotation; it spans time steps 149–278. However, because the trajectory at the end of this movement approaches the starting point of $\rho_1$, a second loop $\rho_2$ emerges.

In addition, during time steps 0--149, t-SNE and UMAP exhibit similar embedding patterns, while {\adahisomap} and Isomap also produce consistent structures. In contrast, TopoAE++ shows substantially greater discrepancies relative to the other methods. Considering the underlying camera motion during this period, the embeddings produced by all methods except TopoAE++ appear reasonable.

\subsection{Network Data}
\label{sec:NetworkData}
\begin{figure}[t]
\centering 
\includegraphics[width=.85\columnwidth]{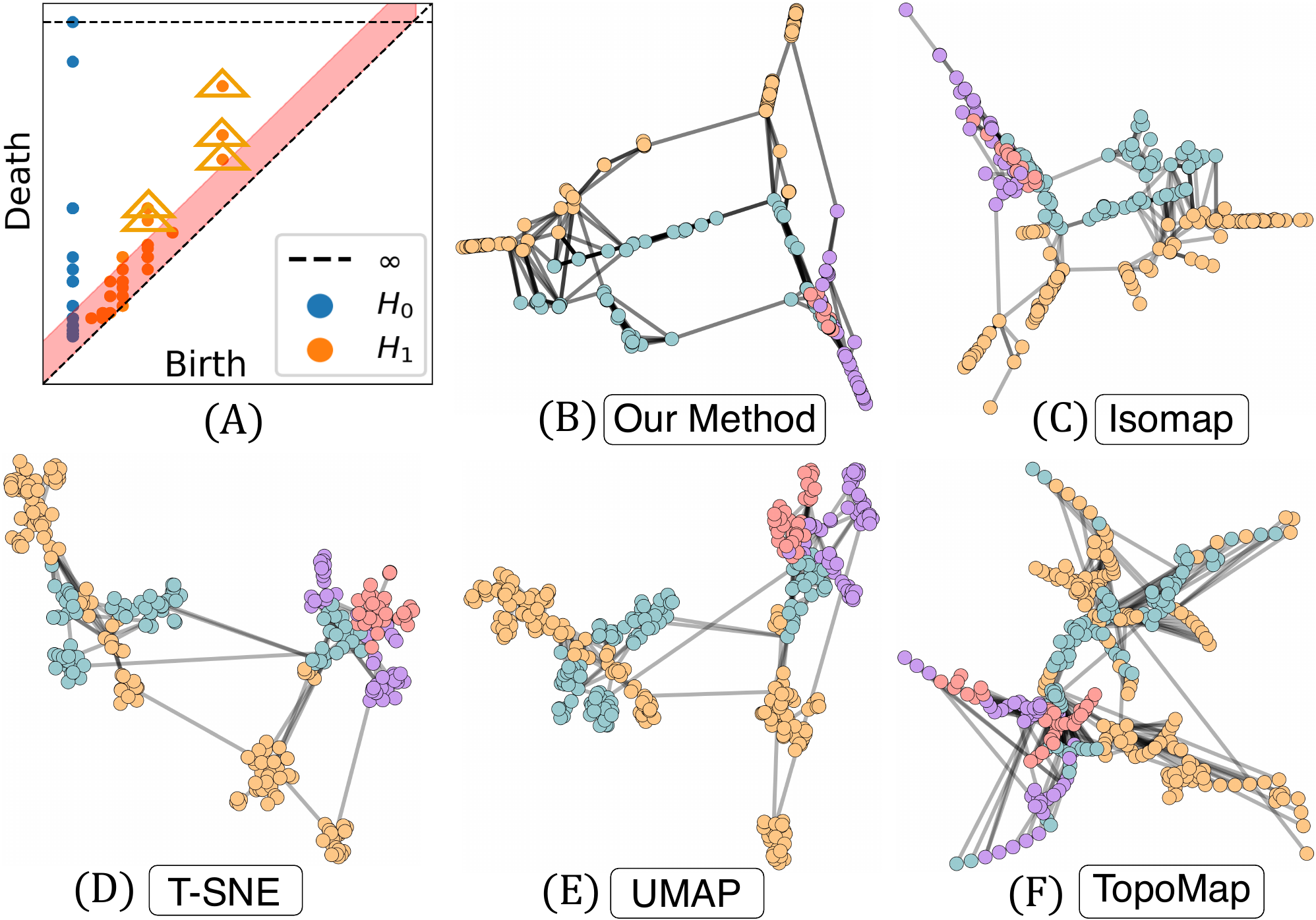}
\caption{Dimensionality reduction of the {\coauthor} dataset. (A) Persistence diagram. (B–F) Embeddings produced by {\adahisomap}, Isomap, t-SNE, UMAP, and TopoMap, respectively, colored by cluster labels.} 
\vspace{-4mm}
\label{fig:coauthor}
\end{figure}

We apply our methods to the \mm{\mathsf{Coauthor}} network dataset, which contains $N=379$ nodes~\cite{Newman2006}. Unlike the previous datasets, the {\coauthor} dataset consists solely of an edge list, where each edge is represented as an ordered pair of nodes (source, destination). Accordingly, we make several minor modifications to the {\adamapper} and {\adahisomap} pipelines. Specifically, we use the shortest-path distance matrix derived from the graph for base-point selection, compute persistence diagrams using the same shortest-path distance matrix, and configure DBSCAN with a precomputed distance metric.
The persistence diagram shown in~\Cref{fig:coauthor}(A) reveals the presence of nontrivial 1D loops in the \coauthor~dataset. Since no prior structural information is available for this dataset, we cluster the data points based on their connections to the four highest-degree nodes.
As shown in~\Cref{fig:coauthor}(B-F), {\adahisomap} provides the clearest visualization of the loopy structures in the resulting 2D embeddings. Examining the clustering distributions further suggests that the embeddings produced by {\adahisomap} and Isomap are more coherent than those of the other methods, as points within the same cluster remain relatively compact. In contrast, the green, red, and yellow clusters are fragmented into multiple disconnected components in the t-SNE, UMAP, and TopoMap embeddings.

Overall, our experiments do not claim that {\adahisomap} universally outperforms all existing DR methods. Rather, they demonstrate that our approach provides a principled mechanism for preserving homological structure in a manner that is both complementary to and competitive with existing approaches.

\vspace{-10pt}
\subsection{Evaluation Metrics}
\label{sec:Eva_Metrics}
We employ three evaluation metrics: one measuring geometric fidelity and two measuring the preservation of 0D and 1D topological features.

Before computing the evaluation metrics, we apply evaluation-stage normalization to all datasets except \coauthor, whose original-space representation is a graph shortest-path distance matrix rather than Euclidean coordinates (see~\Cref{sec:NetworkData}) and thus admits no coordinate-based normalization. No normalization is applied within the DR pipelines themselves. This evaluation-stage normalization provides a common basis for comparing DR methods with different native output scales. Specifically, the input and output of each DR method are normalized independently by centering each coordinate by its mean and dividing all coordinates by the maximum coordinate range. Thus, this procedure corresponds to a uniform isotropic rescaling of each point cloud rather than dimension-wise normalization. It is worth noting that TopoMap's native output attains a near-zero PDW$^0$, which becomes nonzero once the same evaluation-stage normalization used for all other methods is applied.

Hereafter, 
$X$ and $Y$ denote the input and output point sets used for evaluation, normalized except for \coauthor.

\subsubsection{Geometric Evaluation}
To assess geometric preservation, we use the root mean squared error (RMSE), referred to as \emph{metric distortion}, which measures discrepancies between pairwise distances in the input and output spaces. Lower values indicate better geometric preservation. Let $X=\{x_i\}_{i=1}^n$ and $Y=\{y_i\}_{i=1}^n$, where $y_i \in \mathbb{R}^2$. The metric is defined as:
\[
\mathcal{D}(X,Y)=
\sqrt{
\frac{1}{n}
\sum_{1 \leq i < j \leq n}
\left(
\|x_i-x_j\|_2 -
\|y_i-y_j\|_2
\right)^2
}.
\]

\subsubsection{Topological Evaluation}
To evaluate topological faithfulness, we use the \emph{$2$-Wasserstein distance} between the persistence diagrams of the input and output spaces, defined as
\[
    \text{PDW}^j(X,Y)=\left[\adjustlimits \inf_{\eta: \dgm_j(X) \to \dgm_j(Y)} \sum_{x \in \dgm_j(X)} || x - \eta(x) ||_2^2 \right]^{\frac{1}{2}}  
\]
We denote by PDW$^0$ and PDW$^1$ the distances between the 0D and 1D persistence diagrams, respectively.
\section{Conclusion and Discussion}
\label{sec:conclusion}

In this paper, we demonstrate that the preservation of 0D and 1D homological features during dimensionality reduction can be achieved through a homological skeleton extracted using an adaptive Mapper-based framework. Our work also opens several directions for future research.

First, the computational efficiency of both {\adamapper} and {\adahisomap} could be further improved by accelerating the PD-induced segmentation process. Currently, this segmentation relies only on partial information extracted from the persistence diagram. Consequently, the development of algorithms that provide only the persistence information necessary for segmentation---rather than computing the full persistence diagram---could significantly reduce computational cost and benefit both methods.

Second, due to the limitations of the $k$NN graph construction, neither {\adamapper} nor {\adahisomap} can effectively handle datasets containing multiple manifolds. For example, our approach cannot be directly applied to a dataset containing two disconnected faces, such as the example shown in~\cref{fig:face}. One possible direction is to first identify the underlying manifolds and then construct separate $k$NN graphs for each component. The main challenge would then be how to construct a skeleton that captures the global structure of the entire dataset while remaining faithfully representable in low-dimensional space.

Finally, a key remaining challenge is the development of a theory-driven framework for parameter selection in {\adamapper}. Although we provide an automated strategy that performs well empirically, most parameter choices are still guided by heuristic observations and practical experience. Establishing rigorous mathematical foundations for parameter selection would provide stronger theoretical justification and improve the robustness and interpretability of the framework. Recent research offers promising evidence toward this direction~\cite{ChalapathiZhouWang2021,AlvaradoBeltonFischer2025}. 

Overall, this work demonstrates that persistent homology can serve not only as a post hoc analysis tool, but also as an active mechanism for guiding algorithmic structure in dimensionality reduction. We believe this perspective opens new opportunities for topology-informed pipeline design in the analysis and visualization of high-dimensional data.

\section*{Acknowledgment}
This work was supported in part by NSF grant OAC-2313124 and DOE grant DE-SC0021015.

\bibliographystyle{IEEEtran}
\bibliography{AdaHIsomap.bib}

\clearpage
\appendices
\section{Parameter Selection for the DBSCAN Algorithm Based on PD-Induced Segmentation}
\label{sec:clusterParam}

Although the choice of clustering algorithm is not central to our method, we adopt DBSCAN (Density-Based Spatial Clustering of Applications with Noise)~\cite{EsterKriegelSander1996} because of its ability to detect clusters of arbitrary shape and identify noise in spatial datasets. In its original formulation, successful DBSCAN clustering requires suitable choices of the parameters $\epsilon$ (the neighborhood radius) and $minPts$ (the minimum number of points needed to form a dense region), as well as at least one representative point from each cluster. However, in practice, there is no straightforward way to automatically determine appropriate parameter values for all clusters~\cite{EsterKriegelSander1996}.

However, our experiments indicate that, once a suitable value of $\epsilon$ is chosen, the clustering results are relatively insensitive to the choice of $minPts$. To simplify parameter tuning, we therefore fix $minPts = 1$ for all datasets. We then propose an automatic strategy for estimating $\epsilon$ for each point cloud, while still allowing for optional manual adjustment. The procedure is described as follows.

\begin{figure}[h]
\centering
\includegraphics[width=1\columnwidth]{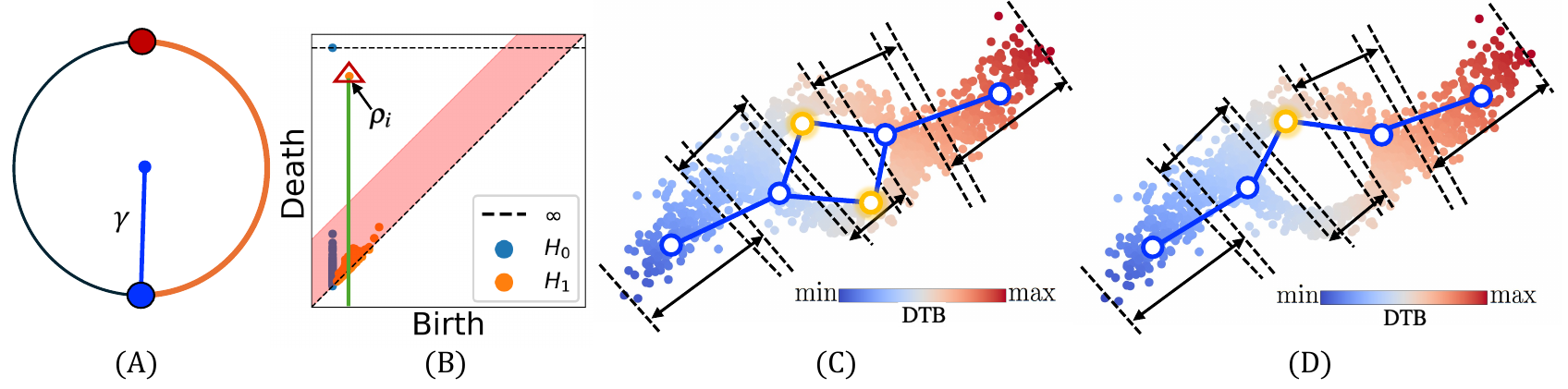}
\vspace{-4mm}
\caption{
Main idea for selecting $\epsilon$.
(A) Intuition behind the choice of $\epsilon$.
(B) Persistence diagram $\mathrm{Dgm}(X)$.
(C) Point cloud $X$ colored by the DTB function $f(X)$, showing the DBSCAN clustering result in the loop region for $\epsilon \approx \rho_i[1]/1.6$, with the identified cluster highlighted in orange.
(D) Point cloud colored by the DTB function, showing the DBSCAN clustering result in the loop region for $\epsilon > \rho_i[1]/1.6$, with the identified cluster highlighted in orange.
}
\vspace{-1mm}
\label{fig:xt}
\end{figure}

\noindent\textbf{Automatic estimation of $\epsilon$.}  
We automatically estimate $\epsilon$ for each point cloud, while still allowing users to override the value if desired. Let $T$ denote the simplification threshold. Among all 1D homological features with persistence greater than $T$, we select the feature with the smallest persistence and denote its death time by $\rho_i[1]$. We then define
$
\epsilon = \frac{\rho_i[1]}{1.6}.
$
The scaling factor $1.6$ is motivated by both geometric intuition and empirical observations. Consider an ideal circular loop of radius $\gamma$ covered by three cubes. To preserve the loop structure during segmentation, the middle cube should contain at least two distinct clusters (see~\cref{fig:xt}(A)), which requires $\epsilon$ to be sufficiently small for density-based clustering to separate points within that region.

We estimate $\gamma$ using the death time $\rho_i[1]$ of the selected loop feature. Experiments on noisy synthetic loops suggest that the relation
$
\gamma \approx \frac{\rho_i[1]}{1.6}
$
provides a reliable balance between preserving the loop structure and avoiding excessive over-segmentation. In particular, this choice of $\epsilon$ typically yields at least two distinct clusters in the central loop region (see~\cref{fig:xt}(C)) without introducing an excessive number of additional clusters.

\section{Dimensionality Reduction Parameters}
\label{sec:parameter}

We present the parameter configurations used for the experimental and evaluation results on 11 datasets. Unless otherwise specified, {\adahisomap} uses the following default settings: simplification threshold $T = 0.35$, overlap percentage $p = 0.2$, minimum number of points $\minPts = 1$, and number of nearest neighbors $NN$. For datasets with $N \geq 1000$, we set $NN = 10$; for datasets with $N < 1000$, we set $NN = 8$; see \Cref{table:parameters}.

\begin{table}[!h]
\captionsetup{justification=raggedright, singlelinecheck=false} 
\scriptsize 
\raggedright 
\setlength{\tabcolsep}{3pt}  
\renewcommand{\arraystretch}{0.9}  
\caption{Isomap, t-SNE and UMAP parameter settings. $NN$ denotes the number of neighbors, $n$ denotes the number of components.}
\label{tab:Drparameters}
\resizebox{\columnwidth}{!}{%
\begin{tabular}{r | c | *{11}{c}}
\toprule
& \rotatebox{90}{\textbf{Parameter}} & \rotatebox{90}{\SH} & \rotatebox{90}{\glasses} & \rotatebox{90}{\fertility} & \rotatebox{90}{$\Octa$} & \rotatebox{90}{\bcsstk} & \rotatebox{90}{\fourelt} &  \rotatebox{90}{\Mice} & \rotatebox{90}{\face} & \rotatebox{90}{\VS} & \rotatebox{90}{\nkx} & \rotatebox{90}{\GIF} \\

\midrule
\multirow{2}{*}{\isomap}
& $n$ & 2 & 2 & 2 & 2 & 2 & 2 & 2 & 2 & 2 & 2 & 2 \\
& $NN$ & 20 & 15 & 15 & 5 & 10 & 15 & 8 & 8 & 6 & 17 & 9 \\
\midrule

\multirow{3}{*}{t-SNE}
& $n$ & 2 & 2 & 2 & 2 & 2 & 2 & 2 & 2 & 2 & 2 & 2 \\
& $perplexity$ & 50 & 50 & 50 & 40 & 30 & 30 & 40 & 40 & 5 & 30 & 30 \\
& $max\_iter$ & 1000 & 5000 & 1000 & 1000 & 1000 & 5000 & 5000 & 1000 & 5000 & 5000 & 1000 \\
\midrule

\multirow{3.5}{*}{UMAP}
& $n$ & 2 & 2 & 2 & 2 & 2 & 2 & 2 & 2 & 2 & 2 & 2 \\
& $NN$ & 35 & 35 & 10 & 30 & 30 & 15 & 25 & 30 & 15 & 40 & 20 \\
& $min\_dist$ & 0.5 & 0.1 & 0.5 & 0.5 & 0.5 & 0.5 & 0.5 & 0.5 & 0.5 & 0.5 & 0.3 \\
\bottomrule

\end{tabular}
}
\label{table:parameters}
\end{table}

We omit {\topomap} and TopoAE++ from the parameter table because, for all datasets, we used the existing TTK (Topology ToolKit)~\cite{TiernyFavelierLevine2017} implementations with default parameters through ParaView 5.13~\cite{AhrensGeveciLaw2005}. Specifically, for {\topomap}, we used the Kruskal MST algorithm with Angular Samples = 2, while for TopoAE++, we used 1000 training epochs with hidden layer dimensions $128$ and $32$. 

For {\isomap} and t-SNE, we used the open-source implementations available in scikit-learn~\cite{PedregosaVaroquauxGramfort2011}, and for UMAP we used the \texttt{umap-learn} implementation~\cite{McinnesHealyMelville2018}, both available in Python. Finally, for {\adamapper}, we used our Python implementation.

\section{Additional Experimental Results} 
\label{sec:AdditionalExperimentalResults}

\begin{figure}[t]
\centering 
\includegraphics[width=\columnwidth]{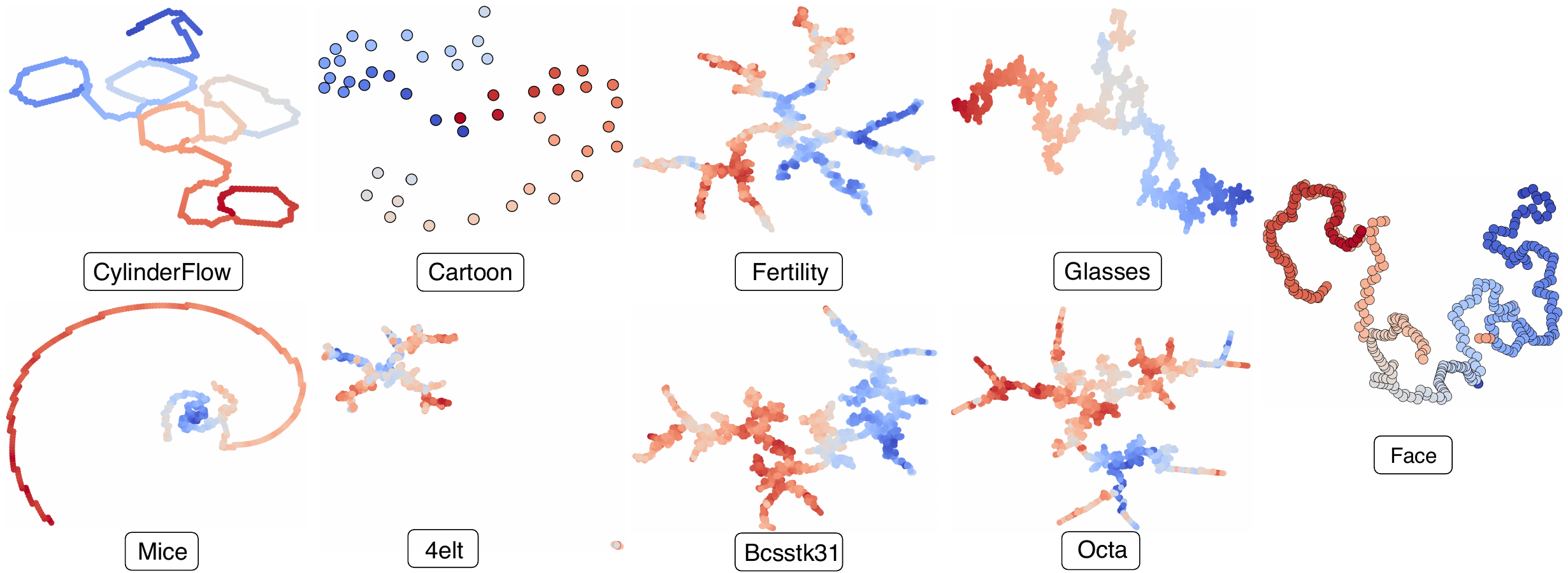}
\caption{TopoMap embeddings for all remaining datasets in Sec.~VI.} 
\vspace{-2mm}
\label{fig:Topomap}
\end{figure}

The TopoMap embeddings are shown in~\cref{fig:Topomap}, and the quantitative analysis is summarized in \Cref{table:parameters}.

As illustrated in~\cref{fig:Topomap}, TopoMap effectively preserves 0D homological features but is less successful at preserving 1D features. \Cref{table:parameters} further shows that, across the 11  datasets, the relative performance of the methods varies depending on the evaluation metric. In terms of RMSE, {\adahisomap} ranks among the top two performers and often outperforms Isomap and TopoMap on datasets such as \glasses, \Mice, \face, and \VS, while UMAP performs particularly well on datasets such as \Octa. 

For WD1, which measures preservation of 1D homology, {\adahisomap} achieves the best result on \fourelt, while TopoAE++ performs best on \fertility. t-SNE and UMAP also achieve top performance on several individual datasets, such as \face~and~\VS. For WD0, which evaluates preservation of 0D homology, TopoMap and UMAP dominate across multiple datasets. In particular, TopoMap performs strongly on \Mice, \glasses, and \nkx, while UMAP achieves the best results on \Octa, \bcsstk, and \fourelt. 

Overall, {\adahisomap} demonstrates robust and well-balanced performance across all evaluation metrics.

\end{document}